%% file: main.tex
\documentclass[aps,floatfix,superscriptaddress,longbibliography,twocolumn,prl]{revtex4-2}
\usepackage{graphicx} 
\usepackage[utf8]{inputenc}
\usepackage{natbib}
\usepackage[normalem]{ulem}
\usepackage{graphicx}
\usepackage{xcolor}
\UseRawInputEncoding

\usepackage[tbtags]{amsmath}
\usepackage[backref=none,
bookmarksnumbered=true,
bookmarks=true,
bookmarksopen=true,
colorlinks=true,
citecolor=blue,
linkcolor=black,
anchorcolor=green,
urlcolor=blue,unicode=false]{hyperref}
\usepackage{amsmath,amssymb}
\usepackage{float}
\usepackage{graphicx}
\usepackage{bbm}
\usepackage{bm}
\usepackage{titlesec}
\usepackage{braket}
\usepackage{todonotes}
\usepackage{siunitx}
\graphicspath{{}}

 \usepackage{commath}
 \usepackage{mathtools}
\usepackage{listings}
 \usepackage{multirow}
\input{kmacros}

\begin{document}
\title{Superfluid weight in disordered flat-band superconductors 
as a competition between localization functionals }
\author{Kry\v{s}tof Kol\'a\v{r}}
\affiliation{Department of Applied Physics, Aalto University School of Science, FI-00076 Aalto, Finland}
\author{Tero T.~Heikkil{\"a}}
\affiliation{Department of Physics and Nanoscience Center, University of Jyv{\"a}skyl{\"a}, FI-40014 University of Jyv\"askyl\"a, Finland}
\author{P{\"a}ivi T{\"o}rm{\"a}}
\email{paivi.torma@aalto.fi}
\affiliation{Department of Applied Physics, Aalto University School of Science, FI-00076 Aalto, Finland}
\date{\today}
\begin{abstract}
According to Anderson's theorem, the gap of a time-reversal symmetric weak-coupling superconductor is unaffected
by non-magnetic disorder. However, the superfluid weight (stiffness) is reduced in the disordered limit by a factor of $\Delta \tau$, a product of the scattering time $\tau$
and the superconducting order parameter $\Delta$.
Here we show that the opposite holds true in flat-band superconductors.
While non-magnetic disorder does reduce the order parameter, we find that its direct effect on superfluid
weight is mostly negligible. We show analytically that the effect of disorder is to lowest order 
given in terms of the difference 
between the intraband and interband parts of the localization functional of impurity wavefunctions, 
finding it to be typically vanishing.
\end{abstract}
\maketitle

Systems with flat bands, where interaction effects dominate, have emerged as an interesting platform to study strongly correlated phenomena,
with recent realizations using the moir\'e effect \cite{jarillo-herreroCaoUnconventionalSuperconductivityMagicangle2018,jarillo-herreroCaoCorrelatedInsulatorBehaviour2018,youngBalentsSuperconductivityStrongCorrelations2020,rubioKennesMoireHeterostructuresCondensedmatter2021,youngAndreiMarvelsMoireMaterials2021,bernevigTormaSuperconductivitySuperfluidityQuantum2022,oliverTanakaSuperfluidStiffnessMagicangle2025,bockrathTianEvidenceDiracFlat2023}, flat surface states \cite{youngZhouSuperconductivityRhombohedralTrilayer2021,youngZhouHalfQuartermetalsRhombohedral2021,juLuFractionalQuantumAnomalous2024} as well as photonic, ultracold atom, and artificial quantum systems~\cite{flachLeykamArtificialFlatBand2018}.
Remarkably, while single-particle transport in clean flat bands is quenched, Cooper pairs are mobile 
\cite{tormaPeottaSuperfluidityTopologicallyNontrivial2015,huberTovmasyanPreformedPairsFlat2018,peottaTormaQuantumMetricEffective2018} and flat
band systems support supercurrent with a finite superfluid weight $\mathcal D_{s}$ determined by the quantum geometry of the flat band eigenstates: for isolated flat bands $\mathcal D_{s} \propto \int d\mathbf k \mathrm{Tr} g(\mathbf k)$, given in terms of a Brillouin zone integral of the trace of the minimal quantum metric $g(\mathbf{k})$~\cite{tormaPeottaSuperfluidityTopologicallyNontrivial2015,tormaHuhtinenRevisitingFlatBand2022}. A salient open question is how disorder changes superfluid weight, and thereby supercurrents and the Meissner effect, in flat bands where its effect could be detrimental due to the lack of a kinetic energy scale. Only a few numerical studies have addressed this question~\cite{hyartLauUniversalSuppressionSuperfluid2022,mondainiLiangDisorderInteractingQuasionedimensional2023,batrouniChanDisorderRobustnessSuperconductivity2025,thuminBouzerarRobustnessFlatBand2025}, indicating that disorder does not necessarily quench superconductivity in flat bands. In this Letter, we 
show using a fully analytical approach,
that the effect of disorder on the flat band superfluid weight can be understood as a competition between intra- and interband localization functionals, revealing the role of quantum geometry in the robustness of flat band superconductivity against disorder.

\begin{figure}[t]
    \includegraphics[width=\columnwidth]{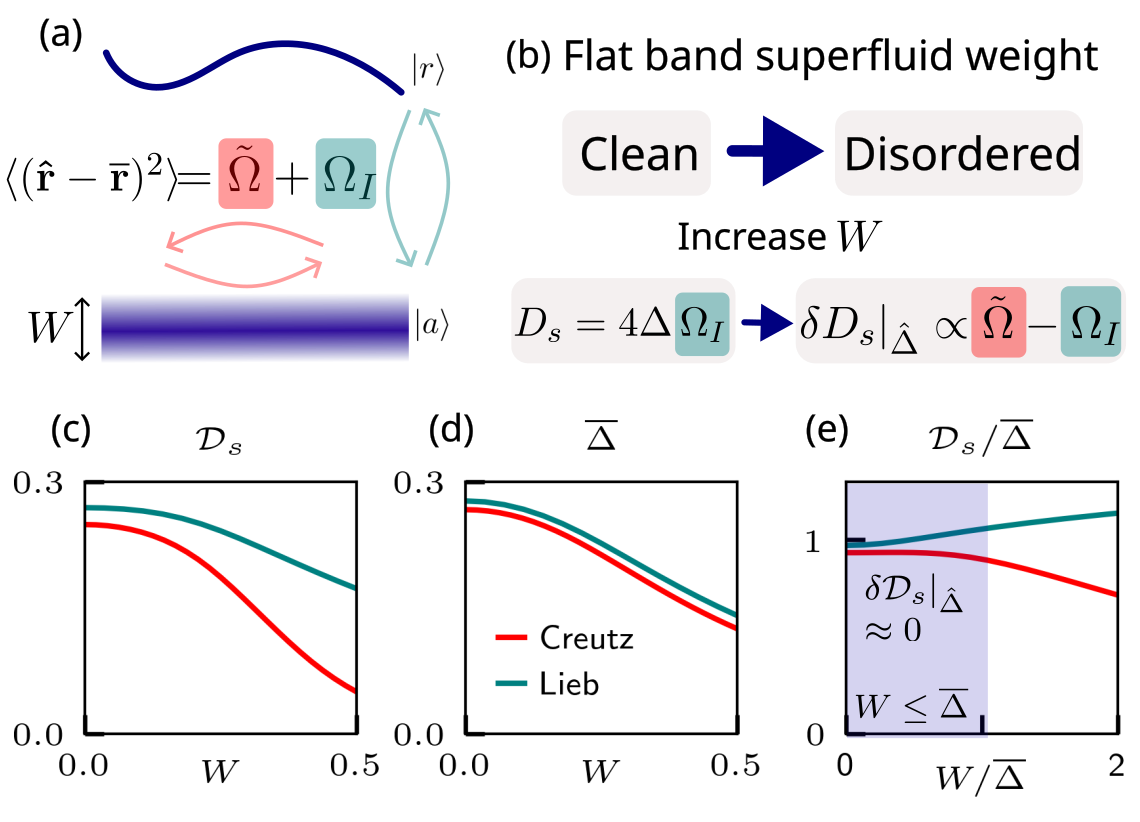}
    \caption{
(a) Schematic illustrating the system considered. Here $W$ is the disorder strength in units of the largest hopping strength, and $|a\rangle$ and $|r\rangle$ are the states in the band of interest and the other bands, respectively.
(b) Illustration of the principal result of this work. 
(c) Mean-field superfluid weight as a function of $W$ for the one-dimensional (1D) Creutz ladder (\colora)
and two-dimensional (2D) Lieb lattice (\colorb). The interaction strength $|U|=1$. See Supplementary Material (SM) \cite{supplement}, Section~\ref{suppl:numericsdetails} for details of the calculation. 
    (d) Same as (c) but for the averaged mean-field gap.
    (e) Ratio of the mean-field superfluid weight to the mean-field gap as a function of $W/\overline \Delta$. Our analytical results predict $\delta \mathcal D_s|_{\hat \Delta} \approx 0$ up to $W/\overline \Delta=1$.
    }
    \label{fig:figone}
\end{figure}

The localization functional (Fig.~\ref{fig:figone}a) of a normalized wavefunction $\ket{A}$ can be defined as
\begin{equation}
\label{eq:locfunctional}
\braket{A|(\posop-\overline{\mathbf r})^2| A} = \Omega_I(\ket{A}) + \tilde \Omega(\ket{A}),
\end{equation}
where the interband and intraband parts are, respectively
\begin{eqnarray}
\Omega_I(\ket{A}) =\braket{A|\posop [1-\projfb] \posop |A}\label{eq:locfunctionalI} \\
\tilde \Omega(\ket{A}) = \braket{A|\posop \projfb \posop | A}-|\overline{\mathbf r}|^2 \label{eq:locfunctionalII},
\end{eqnarray}
where $\projfb$ projects on the flat band.  
In the theory of Wannier functions, $\Omega_I$ is called the gauge-invariant contribution \cite{vanderbiltMarzariMaximallyLocalizedGeneralized1997,yangYuQuantumGeometryQuantum2024}, as it is the same for any choice of the Wannier function. It
is equal to the integrated quantum metric, $\Omega_I(\ket{\text{WF}}) = \frac{1}{(2\pi)^d}\int d\mathbf k \mathrm{Tr} g(\mathbf k)$, and is therefore related to the superfluid weight of an isolated flat band superconductor.

Here, we show that the disorder-induced 
change in the superfluid weight is proportional to the difference of 
the two localization functionals: $\delta \mathcal D_s \propto [\tilde \Omega(\ket{\boundstate}) - \Omega_I(\ket{\boundstate}]$, defined for the bound state wavefunction $\ket{\boundstate}$ of a single impurity. Physically, the
intraband increase proportional to $\tilde \Omega$ arises due to disorder-enabled current within the flat band.
The interband reduction $\Omega_I$ reflects a disorder-driven decrease in the interband contribution $\mathcal D_s^{inter}$ of the total superfluid weight. 
Importantly, since the two parts of the localization functional for an impurity are typically of similar magnitude (as will be discussed below), the change is generally very small, implying that the superfluid weight of a flat band superconductor is robust to disorder. 
 
We decompose the total change in the superfluid weight due to disorder, $\delta \mathcal D_s$, into two effects: (i) the reduction of the elements of the order parameter matrix (in the orbital basis) $\gapmat$,
(ii) the effect of the change in the eigenstate structure while keeping $\gapmat$ constant:
\begin{equation}
\label{eq:dsdecomposition}
\delta \mathcal D_s =  \frac{\partial\mathcal D_s}{\partial \gapmat} \delta \gapmat+  \delta \mathcal D_s|_{\gapmat}.
\end{equation}
While the order parameter does get reduced by disorder, see e.g. Ref.~\cite{trivediGhosalRoleSpatialAmplitude1998}, effects arising from the change of the wavefunctions only are  
inherent in the second term. 
By focusing on the second term of Eq.~\eqref{eq:dsdecomposition}, we are able to study the effect of disorder analytically,
revealing the role of localization functionals in the robustness of flat band superconductivity. Our main result is summarized in Fig~\ref{fig:figone}b. 
We support our main, analytically-derived, result 
with unconstrained self-consistent mean-field calculations, shown in Figs~\ref{fig:figone}c,d,e.
While the calculated superfluid weight shown in Fig.~\ref{fig:figone}c does decrease with increasing disorder strength $W$, this decrease is essentially
the result of the reduction of the gap, shown in Fig.~\ref{fig:figone}d.
Dividing the superfluid weight by the orbital- and disorder-averaged order parameter $\overline \Delta$ provides a proxy for
$\delta \mathcal D_s|_{\gapmat}$, and we find that this quantity is robust to disorder up to $W \approx \overline \Delta$, cf. Fig.~\ref{fig:figone}e.

{\em Model.---}
We consider an $\norb$-orbital model possessing for a single spin species the single particle Hamiltonian $\hclean =\sum_{\mathbf k} \ket{\mathbf k}\bra{\mathbf k}\otimes \hkin$, which is at each momentum $\mathbf k$ 
diagonalized by $\norb$-dimensional spinors
$\hkin \ket{\psi_m(\mathbf k)} =\epsilon_{m,\mathbf k} \ket{\psi(m,\mathbf k)}$
with energies $\epsilon_{m,\mathbf k}$ (which include the chemical potential),
and the full wavefunctions are obtained as 
$\ket{\Psi(m,\mathbf k)} = \ket{\mathbf k} \otimes \ket{\psi_m(\mathbf k)} $.
We separate the bands into an active band near to the Fermi energy (with $m=m_{FB}$) and remote bands,
using the index $a \in FB$ to label the active states and $r \in R$ to label the remote states, see Fig.~\ref{fig:figone}a.
Without disorder, these are just the band and momentum labels.

Next, we introduce a disorder term $\hdis$ corresponding to uncorrelated on-site disorder
\begin{equation}
\label{eq:disorderham}
\hdis =  \projfb \sum_{\mathbf r, \slind} \epsilon_{\mathbf r,\slind} \ketbra{\mathbf r,\slind} \projfb,
    \end{equation}
where $\ket{\mathbf r,\alpha}$ is localized on site $\mathbf r$ and orbital $\alpha$, and
where we assume the on-site energies $\epsilon_{\mathbf r,\slind}$ to be independent and identically distributed with the variance
$\disav{ \epsilon_{\mathbf r,\slind}^2 } = W^2$.
Note that flat band eigenstates are localized even without disorder and remain localized when generic disorder is added \cite{flachCadezMetalinsulatorTransitionInfinitesimally2021}.
Importantly, we project the disorder term onto the active band subspace,
which preserves the decomposition into single particle bands of $\hkin$.
The eigenstates of the combined Hamiltonian $\htot = \hclean + \hdis $ differ from $\hclean$ only in the active band and
each active band eigenstate can be written as a superposition of the clean eigenstates:
\begin{equation}
\ket{a} = \int d \mathbf k \frac{f_a(\mathbf k)}{(2\pi)^d}  \ket{\Psi(m_{FB},\mathbf k)}  \label{eq:eigenstatesuperposition},
\end{equation}
where $f_a(\mathbf k)=\braket{a|\Psi(m_{FB},\mathbf k)}$ are expansion coefficients for the state $a$.

Next, we turn to the paramagnetic current operator,
which enters the calculation of superfluid weight.
Defined as
$\currop = i\left[\posop, \htot \right]$, it
vanishes within the flat band subspace for a clean system.
Importantly, 
since $\projfb\left[\posop,\hdis\right]\projfb \neq 0 $, projected disorder enables current in a flat band with quantum geometry.
Note that the same result holds for unprojected disorder, where flat band current is enabled by an
admixture of remote band states \cite{lawChauDisorderinducedDelocalizationFlatband2024}. As we show in the SM~\cite{supplement}, Section~\ref{suppl:projectionworks}, the current in this case is equal to the projected case to leading order in $W/E_G$, where $E_G$ is the band gap between the active and remote bands. 
We consider the single-particle Hamiltonian $\hat{H}$ paired with its time-reversed partner in an s-wave superconductor
with uniform pairing, using the mean-field Hamiltonian
\begin{equation}
\label{eq:hbdgdis}
 \hbdg= \begin{pmatrix}\hclean + \hdis  & \gapmat \\ \gapmat&-\hclean - \hdis\end{pmatrix},
\end{equation}
where the uniform gap operator reads
\begin{equation}
\gapmat = \sum_{\mathbf r, \slind} \gap \ket{\mathbf r,\alpha}\bra{\mathbf r,\alpha}, \label{eq:UPA}
\end{equation}
being proportional to the identity matrix.
This Hamiltonian is solved explicitly using the single particle eigenstates $\ket{s}$ of $\htot =\hclean + \hdis$ as
\begin{eqnarray}
\ket{s,\pm}   = \ket{\pm_s} \otimes \ket{s}\\
\ket{+_s} = \begin{pmatrix}u_s \\ v_s  \end{pmatrix} ; \ket{-_s} = \begin{pmatrix}v_s \\ -u_s  \end{pmatrix} 
\end{eqnarray}
having energies $E_{s,\pm} = \pm \sqrt{\gap^2+(\epsilon_{s})^2} = \pm E_{s}$, and where $u_s,v_s$ are solutions to the reduced Bogoliubov de Gennes problem
\begin{equation}
\label{eq:hbdgdisreduced}
\begin{pmatrix}  \epsilon_\disindex   & \gap \\ \gap&-\epsilon_\disindex\end{pmatrix}
\begin{pmatrix}u_\disindex  \\ v_\disindex  \end{pmatrix}  = E_{\disindex}
\begin{pmatrix}u_\disindex  \\ v_\disindex  \end{pmatrix}.
\end{equation}
In the clean case, the uniform pairing condition (same $\Delta$ at all orbitals where it is non-zero)~\cite{huberTovmasyanEffectiveTheoryEmergent2016} holds
for attractive Hubbard models possessing symmetries that connect different orbitals~\cite{tormaHuhtinenRevisitingFlatBand2022,bernevigHerzog-ArbeitmanManyBodySuperconductivityTopological2022}. 
With disorder, pairing is never strictly uniform. The uniform pairing approximation Eq.~\eqref{eq:UPA} is instead used
to understand at the qualitative level the effect of the changes in wavefunctions by disorder,
i.e., the second term in Eq.~\eqref{eq:dsdecomposition}.

{\em Disordered superfluid weight.---}
Adapting the linear response approach of \cite{tormaLiangBandGeometryBerry2017}, the trace of the zero temperature superfluid weight within the uniform pairing approximation becomes
\begin{equation}
\label{eq:sfweightgeneral}
\mathcal D_{s}=\frac{1}{\nuc}\sum_{s_1,s_2} \frac{2}{E_{s_1}+E_{s_2}} 
\frac{\Delta}{E_{s_1}} \frac{\Delta}{E_{s_2}} |\braket{s_1|\currop|s_2}|^2 ,
\end{equation}
where the sum is over all pairs of \textit{single-particle} eigenstates $s_1,s_2$ of $\htot$, and $\nuc$ is the number of unit cells. 
Furthermore, the diamagnetic contribution is rewritten in terms of the paramagnetic current operator.

We use the minimal quantum metric orbital positions, for which the contribution due to the response of the order parameter to the vector potential vanishes in the clean case~\cite{tormaHuhtinenRevisitingFlatBand2022}. While our analytical approach neglects the order parameter response even in the disordered case, we verify our results using full numerical mean-field calculations below.

We separate the superfluid weight into intraband 
($s_1, s_2 \in FB$)
and interband terms ($s_1 \in R$ or $s_2 \in R$) and disregard the small contribution with $s_1,s_2 \in R$.
The interband, or geometric, term arises due to the momentum dependence of the active band wavefunctions.
Assuming the isolated band limit ($|E_r|\gg |E_a|$), it can be approximated as  
\begin{eqnarray}
\mathcal D^{inter}_{s}& =&\frac{1}{\nuc}\sum_{r \in R,a \in FB} \frac{4\Delta^2}{E_{r}+E_{a}} 
\frac{|\braket{r|[\posop,H]|a}|^2}{E_{r} E_{a}}  \\
&=&\frac{1}{\nuc}\sum_{r \in R,a \in FB} \frac{4\Delta^2(\epsilon_r-\epsilon_a)^2}{E_{r}+E_{a}} 
\frac{|\braket{r|\posop|a}|^2}{E_{r} E_{a}}  \\
 &\approx& \frac{1}{\nuc} \sum_{a \in FB}  \frac{4\Delta^2}{E_{a}}
 \braket{a|\posop(1-\projfb)\posop|a}  \\
\mathcal D^{inter}_{s}&=& \frac{1}{\nuc} \sum_{a \in FB}  \frac{4\Delta^2}{E_{a}}
 \Omega_I(\ket{a}) , \label{eq:diffsfweightinter}
\end{eqnarray}
giving a result in terms of $\Omega_I$, the interband part of the localization functional in Eq.~\eqref{eq:locfunctional}.

The form of the interband contribution in Eq.~(\ref{eq:diffsfweightinter}) directly implies some amount of robustness to disorder.
The localization term $\Omega_I(\ket{a}) = \int d \mathbf k \frac{|f_a(\mathbf k)|^2}{(2\pi)^d} \mathrm{Tr}\, g(\mathbf k) $ is an average of the quantum metric with the weights given by  Eq.~\eqref{eq:eigenstatesuperposition}. 
Ignoring the factor of $E_a$ in the denominator of Eq.~(\ref{eq:diffsfweightinter}), the sum over $a$ then becomes the integrated quantum metric, independent of the nature of the disordered eigenstates.
For a dispersive band, the density of states is largely unchanged by disorder \cite{andersonAndersonTheoryDirtySuperconductors1959}; disorder mixes eigenstates with energies within a shell of width $1/\tau$. Intuitively, thus,
 if the localization functional does not vary strongly for eigenstates within this energy range, the interband contribution [Eq.~\eqref{eq:diffsfweightinter}] is largely unaffected by disorder. 
For a flat band, disorder alters the $1/E_a$ term by a contribution proportional to $\frac{W^2}{\Delta^2}$,
which we show below to be typically compensated by an increase in the intraband superfluid weight.

{\em Disordered superfluid weight in a flat band.---}
We now concentrate on a flat band at zero energy, at which point it is half-filled
(see SM~\cite{supplement}, Section~\ref{suppl:numericsdetails} for a discussion of filling). Concretely we take $\epsilon_{m_{FB},\mathbf k}=0$ in the clean case. For simplicity, we furthermore assume that the flat band orbitals are related by symmetry, referring to the SM~\cite{supplement}, Section~\ref{suppl:generalizerankone} for a general treatment.
Without disorder, only the interband term contributes with $E_a=\Delta$ \cite{tormaPeottaSuperfluidityTopologicallyNontrivial2015}, and Eq.~\eqref{eq:diffsfweightinter}
evaluates to $\mathcal D_{s}\approx  4\Delta\Omega_{I}$.
Evaluating Eq.~\eqref{eq:sfweightgeneral} in the weak disorder limit,
we decompose the superfluid weight change as 
\begin{equation}
\label{eq:diffsfweight}
\left.\delta \mathcal D_{s}\right|_{\gapmat}=\delta \mathcal D^{intra}_{s}+\delta \mathcal D^{inter}_{s}
\end{equation}
where the first term is the enhanced superfluid weight due to disorder-induced intraband current,
while the second term is the reduction of the interband weight due to disorder.

We first consider the effect of a single random impurity, obtained by restricting the sum in
Eq.~\eqref{eq:disorderham} to a single term only with $\mathbf r=\mathbf r_{\boundstate}$ and $\slind= \slind_{\boundstate}$.
Such an impurity produces precisely one bound state (BS) (i.e., it is a  rank-1 impurity) on the flat band~\cite{sternQueirozRingStatesTopological2024},
given as $\ket{\boundstate} =\sqrt{N_f} \projfb\ket{\mathbf r_{\boundstate}, \slind_{\boundstate}},$
with the energy $\epsilon_{\boundstate} =\frac{\epsilon_{\mathbf r, \slind}}{ N_{f}}$, where $N_f$ is the number of flat band orbitals. Note that here we use that different flat band orbitals are symmetry related. 
By expanding $E_{\boundstate}$ in Eq.~\eqref{eq:diffsfweightinter} to lowest order in $W/\Delta$,
we obtain that the interband superfluid weight is to lowest order in $W/\Delta$ reduced by an amount
$\delta \mathcal D^{\text{inter}}_{s}\approx -\frac{2}{\nuc} \frac{{W}^2}{\Delta N_f^2}  \Omega_I(\ket{\boundstate})$ (see SM~\cite{supplement}, Section~\ref{suppl:interband}),
proportional to the interband localization functional of the bound state wavefunction. Note that this functional
is equal to the integrated quantum metric 
when the flat band orbitals are related by a symmetry (see SM~\cite{supplement}, Section~~\ref{suppl:locfuncdetails}).

The impurity additionally enables current on the flat band, leading to a nonzero and positive intraband contribution to
the superfluid weight: 
\begin{eqnarray}
\delta \mathcal D^{\text{intra}}_{s}&\approx& \frac{2}{\nuc}\sum_{s_1 \in FB,s_2 \in FB}  
\frac{1}{2\Delta}|\braket{s_1|\currop|s_2}|^2 \\
&=& \frac{1}{\nuc}\frac{1}{\Delta}  \mathrm{Tr}[\projfb \currop \projfb \currop]\\
&=&\frac{1}{\nuc}\frac{1}{\Delta}  \mathrm{Tr}\left\{\projfb [\posop,\hdis] \projfb [\hdis,\posop]\right\}\\
&=&\frac{2}{\nuc}\frac{1}{N_f^2}\frac{ W^2}{\Delta}  \tilde \Omega(\ket{\boundstate}) \label{eq:impurityintrabandsfweight},
\end{eqnarray}
so that the total change in superfluid weight for a single impurity, 
\begin{equation}
\delta \mathcal D_{s}^{\text{Impurity}} = \frac{2W^2}{\nuc N_f^2 \Delta}
\left[\tilde \Omega\left(\ket{\boundstate}\right)-\Omega_I\left(\ket{\boundstate}\right) \right] ,
\end{equation}
is a competition between the two parts of the localization functional [Eq.~\eqref{eq:locfunctional}] of the impurity bound state.

We proceed to consider the effect of the full disorder Hamiltonian in Eq.~\eqref{eq:disorderham}.
Due to the independence of the impurity potentials on different sites, the intraband increase in superfluid weight is
equal to $\nuc N_{f}$ times the single impurity contribution (see SM~\cite{supplement}, Section~\ref{suppl:intraband}): 
$\delta \mathcal D^{intra}_{s}\approx 2\frac{ W^2}{N_f \Delta}  \tilde \Omega(\ket{\boundstate})$. 
The interband term is given in terms of exact flat band eigenstates as in
Eq.~\eqref{eq:diffsfweightinter}, and its change is to lowest order in $W/\Delta$ given as 
$\delta \mathcal D^{inter}_{s}\approx -\frac{2}{\nuc}\sum_{a \in FB} \frac{\epsilon_{a}^2}{\Delta}  \Omega_I(\ket{a})=
-\frac{2}{\nuc \Delta}\Tr\left[\hdis^2 \posop\left(1-\projfb\right)\posop   \right]=-\frac{2W^2}{ N_f \Delta}\Omega_I(\ket{\boundstate}) $, arising like the intraband part as  $\nuc N_f$ times the single impurity contribution.
The total change 
\begin{equation}
\label{eq:deltadsfull}
\left.\delta \mathcal D_{s}\right|_{\gapmat}  = 
2\frac{ W^2}{N_f \Delta}  \left[\tilde \Omega(\ket{\boundstate}) -\Omega_I(\ket{\boundstate})     \right]
\end{equation}
is thus a competition of intraband and interband localization functionals of the bound state, just like for a single impurity. Crucially, we find below that the two parts of the localization functional for bound states are typically of similar magnitude. Consequently, we expect the total superfluid weight of flat bands to be largely unaffected by disorder.
The above formula is the principal result of this work.

{\em Examples.---}
We now turn to specific examples to test the approximate analytical results by numerics. We list in Table~\ref{tab:table1} the relevant quantities that, based on the above results, determine the superfluid weight change for the one-dimensional Creutz ladder and two-dimensional Lieb lattice. The Creutz ladder flat band is spanned by compact Wannier states, and the inter- and intraband localization functionals can be analytically determined to be equal, implying $\delta \mathcal D_s = 0$.
The Lieb lattice has three orbitals in a unit cell, but the flat band is supported only by $N_f=2$ of them.
Taking the hopping parameters $t_1=1,t_2=0.8$, we find numerically that 
both functionals are equal, and we expect $\delta \mathcal D_s = 0$ as well;
intriguingly, we find that the difference of the two functionals for the Lieb lattice depends on the orbital positions, and vanishes at the orbital positions that minimize the integrated quantum metric $\Omega_I$ (see SM~\cite{supplement}, Section~\ref{suppl:locfuncdetails}). 

\begin{table}
\centering
\begin{tabular}{|c|c|c|c|}
 \hline
 \raisebox{-0.2ex}{System} & \raisebox{-0.5ex}{$\Omega_I(\ket{\boundstate})$  }&\raisebox{-0.5ex}{$\tilde \Omega(\ket{\boundstate})$}\\ [0.5ex]
 \hline
 Creutz ladder (1D)   &$\frac{1}{4}$ &$\frac{1}{4}$  \\[0.2ex]
 \hline
 Lieb (2D)   &  $0.35$ &$0.35$ \\[0.2ex]
 \hline
\end{tabular}
\caption{Values of $\Omega_I$, $\tilde \Omega$ for a single impurity bound state $\ket{\boundstate}$
for the Creutz ladder and the Lieb lattice. Lattice constant is set to $a=1$. For parameters, see the text.}
\label{tab:table1}
\end{table}

\begin{figure}[t]
    \centering
    \includegraphics[width=\columnwidth]{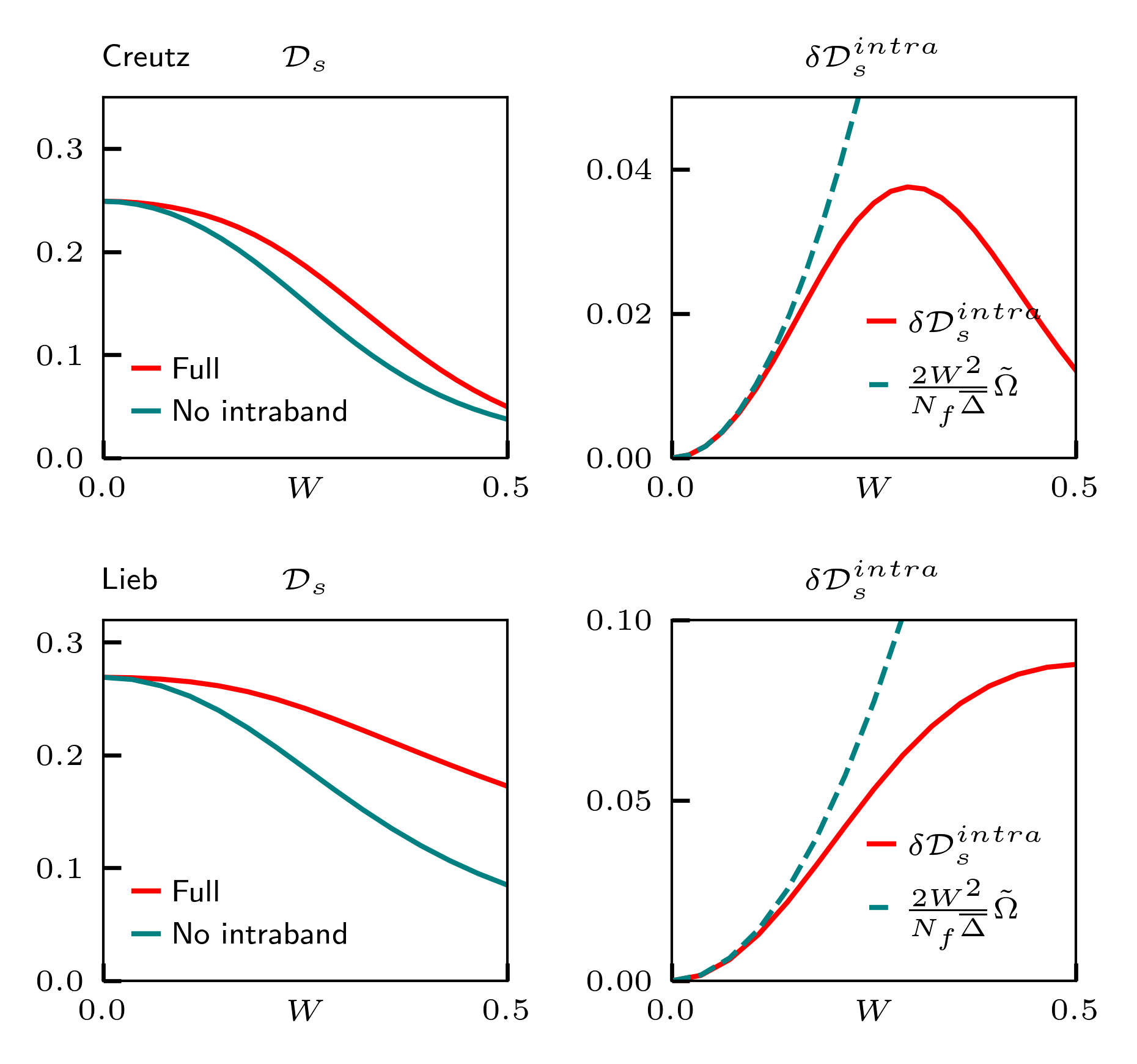}
    \caption{
(a) Superfluid weight as a function of disorder for the 1D Creutz ladder. The full weight is shown in \colora, and the value without intraband terms in \colorb. 
    (b) The intraband contribution to the superfluid weight (full \colora), obtained as the difference between the full and the value without the intraband terms.
    The analytical expression is shown in dashed \colorb.
    (c) Same as (a) but for the 2D Lieb lattice.
    (d) Same as (b) but for the 2D Lieb lattice.
The interaction used was $U=1$ in units of hopping. A gap was opened in the Lieb lattice by asymmetric hoppings. See SM~\cite{supplement}, Section~\ref{suppl:numericsdetails} for parameters and numerical procedures.
    }
    \label{fig:figtwo}
\end{figure}

We verify these analytical predictions within a self-consistent superfluid weight calculation where
pairing is mediated by an on-site Hubbard attraction of strength $|U|=1$ \cite{tormaPeottaSuperfluidityTopologicallyNontrivial2015,tormaLiangBandGeometryBerry2017,tormaJulkuGeometricOriginSuperfluidity2016}
decoupled on every site in the pairing channel
$Uc^\dagger_{\mathbf r, \slind,\uparrow}c^\dagger_{\mathbf r, \slind,\downarrow} 
c^{\phantom{\dagger}}_{\mathbf r, \slind,\downarrow}c^{\phantom{\dagger}}_{\mathbf r, \slind,\uparrow} \to 
\left[ c^\dagger_{\mathbf r, \slind,\uparrow}c^\dagger_{\mathbf r, \slind,\downarrow} 
\Delta_{\mathbf r,\slind} +\hc\right]$,
where the mean-field order parameter
$\Delta_{\mathbf r,\slind} = U\mfav{ c^{\phantom{\dagger}}_{\mathbf r, \slind,\downarrow}c^{\phantom{\dagger}}_{\mathbf r, \slind,\uparrow} }$.
We do not include the Hartree term from the Hubbard attraction, which does not affect the result qualitatively.
Details are in SM~\cite{supplement}, Section~\ref{suppl:numericsdetails}.

While the Creutz ladder is a rather idealized model, in the Lieb lattice one can with one tuning parameter (non-uniform hopping) open a gap between the flat band and other bands. We choose this gap to be small, $E_G \approx \Delta$, violating the isolated band limit. Furthermore, the assumption of uniform pairing at all orbitals is not valid even in the clean case~\footnote{Our assumption is distinct from so-called uniform pairing condition~\cite{huberTovmasyanEffectiveTheoryEmergent2016} which assumes uniform pairing on those orbitals where it is non-zero, and is valid in the Lieb lattice with uniform hoppings.}. This allows us to test the sensitivity of our theory to the assumptions used to obtain the analytical results above.
We calculate the superfluid weight as the total derivative of the grand potential 
$\ds =\frac{1}{\nuc}\sum_i \frac{d \Omega^{\text{Grand}}(\mathbf A)}{d A_i dA_i}$
with respect to an imposed vector potential $\mathbf A$. 
The use of total derivatives is crucial to make the superfluid weight orbital-position (orbital-embedding) independent \cite{rudnerSimonContrastingLatticeGeometry2020,tormaHuhtinenRevisitingFlatBand2022,peottaPeottaSuperconductivityGeneralizedRandom2022,peottaTamGeometryindependentSuperfluidWeight2024}. 

We show the resulting numerical disorder-averaged $\mathcal D_{s}$ in Fig.~\ref{fig:figone}c. We see that $D_s$ decreases with increasing disorder.
However, the total change (Eq.~\eqref{eq:dsdecomposition}) contains both the effects that are the focus of this work, as well as the simultaneous effect of a decrease in the gap.
As shown in Fig.~\ref{fig:figone}d, the gap indeed decreases with increasing disorder, as
is typical for strong coupling superconductivity \cite{trivediGhosalRoleSpatialAmplitude1998,litakMoradianSuperconductingAlloysWeak2000}.
To isolate the geometric effects, we plot in Fig.~\ref{fig:figone}e the normalized superfluid weight $\mathcal D_{s}/{\overline \Delta}$, finding that for both examples it is nearly constant up to $W/{\overline\Delta} \approx 1$.
For the Lieb lattice, we find that, remarkably, the normalized superfluid weight experiences a slight increase with disorder.
This behavior can be rationalized by the violation of the isolated band and uniform pairing assumptions that are made in the above derivation of the interband 
change $\delta \mathcal D_s^{\text{Inter}}$, leading to an overestimate of the interband decrease compared to the numerical result.

The two contributions to the disorder-induced
change of superfluid weight can be isolated by modifying the current operators.
In particular, we remove the intraband contribution to the superfluid weight by subtracting the intraband current
operator from the Hamiltonian at finite vector potential, see SM~\cite{supplement}, Section~\ref{suppl:removeintradetails} for details.
As shown in Fig.~\ref{fig:figtwo}a for the Creutz ladder and Fig.~\ref{fig:figtwo}c for the Lieb lattice,
the superfluid weight without the intraband contribution (\colorb) decreases faster than the full weight (\colora) upon increasing disorder.  The difference of the two isolates the intraband contribution, shown in 
\colora{} in Figs.~\ref{fig:figtwo}b,d. It increases quadratically for small disorder, and
matches well with the theoretical expectation of the first term on the right hand side of Eq.~\eqref{eq:deltadsfull}, shown in dashed \colorb{}.  We note that since for the 2D Lieb lattice, 
our self-consistent calculations are restricted to small system sizes ($\nuc=196$),
we use the same small system to obtain the expected value of $\tilde \Omega\left(\ket{\boundstate}\right)$, which is somewhat smaller than the thermodynamic limit answer given in Table~\ref{tab:table1}.
We give a parameter dependence of these numerical results in the SM~\cite{supplement}, Section~\ref{suppl:extendeddata}.

{\em Discussion.---}
We have analytically studied the effect of disorder on the superfluid weight in flat-band superconductors,
uncovering a competition between interband and intraband localization functionals of the relevant disordered wavefunctions. Within the uniform pairing approximation, our result predicts that the ratio $\mathcal D_s/\overline \Delta$ for flat band superconductors is independent of disorder strength up to $W \sim \Delta$, a prediction that can be checked experimentally for different devices having different disorder strengths. 
We have confirmed numerically within the Lieb lattice model that the qualitative results are robust even when the isolated flat band and uniform pairing conditions are not fulfilled. 

We expect our results to apply generically beyond the specific models considered here. 
Crucially, the two parts of the localization functional for bound states are commonly of similar size.
For instance, in one dimension, Wannier functions diagonalizing $\projfb \posop \projfb$ can be found.  
Writing the bound state wavefunctions in terms of such Wannier functions, it can be shown that 
 $\tilde \Omega \approx \Omega_I$.
In the zeroth Landau level, the guiding center algebra \cite{shankarMurthyHamiltonianTheoriesFractional2003} guarantees that the size of any state satisfies $\tilde \Omega \geq1$ \cite{devereauxClaassenPositionMomentumDualityFractional2015,khalafLiConstraintsRealSpace2024}. The interband functional $\Omega_I=1$ -- a manifestation of the ideal quantum geometry of Landau levels \cite{royRoyBandGeometryFractional2014,yangWangExactLandauLevel2021}; taken together, this means that the effect of disorder on the superfluid weight due to the wavefunction change is either zero or an increase. 
Our analytical approach can be trivially extended to other rank-1 disorder, for instance Wannier disorder, see SM~\cite{supplement}, Section~\ref{suppl:generalizerankone}.

Our results also apply to dispersive superconductors. 
In particular, Eq.~\eqref{eq:diffsfweightinter} shows that the interband contribution to superfluid weight
is given in terms of the interband part of the localization functional of the disordered eigenstates of the active band, and we expect it to be robust to disorder. For dispersive systems, however, this contribution
exists alongside the conventional superfluid weight. 
For a strongly disordered dispersive superconductor, the intraband superfluid weight is proportional to $\Delta \tau$, so that in this limit the interband (geometric) part of the superfluid weight could become dominant. 
Intriguingly, intraband superfluid weight also receives a contribution proportional to the intraband localization functional $\tilde \Omega$ in the disordered dispersive case. Our results show that localization functionals capture the quantum geometric effects of disorder on supercurrent. We envision they could be used also for understanding finite-frequency dissipative and supercurrent transport, as well as non-equilibrium effects \cite{narangAroraQuantumGeometryInduced2025}, both in flat and dispersive bands.  \\

\begin{acknowledgments}
We thank Eeli Lamponen for help with the numerical calculation and Sebastiano Peotta for useful comments on the manuscript. 
We thank Guodong Jiang, Pauli Virtanen, Tim Kokkeler, Junze Deng, and Anushree Datta for useful discussions. 
This work was supported by the Research Council of Finland under project numbers 339313 and 354735, by Jane and Aatos Erkko Foundation, Keele Foundation, Magnus Ehrnrooth Foundation, and a collaboration between The Kavli Foundation, Klaus Tschira Stiftung, and Kevin Wells, as part of the SuperC collaboration, and by a grant from the Simons Foundation (SFI-MPS-NFS-00006741-12, P.T.) in the Simons Collaboration on New Frontiers in Superconductivity. It is also part of the Finnish Centre of Excellence in Quantum Materials (QMAT).
\end{acknowledgments}

\IfFileExists{klibrary.bib}{\bibliography{klibrary.bib}}{\IfFileExists{../klibrary.bib}{\bibliography{../klibrary.bib}}{}}

\begin{widetext}

\pagebreak

\onecolumngrid

\setcounter{secnumdepth}{2}
\begin{center}
    \large \bfseries Supplementary material
\end{center}

\renewcommand{\theequation}{S\arabic{equation}}
\setcounter{equation}{0}

\renewcommand{\thefigure}{S\arabic{figure}}
\setcounter{figure}{0}
\section{Details on the analytical calculations}
\subsection{Change of the interband term with disorder}
\label{suppl:interband}
To derive the interband change in superfluid weight, we make the following two assumptions.
First, we assume that the single-particle energies are proportional to the disorder strength: $\epsilon_a \propto W$.
Second, we assume that $W\ll \Delta$, allowing us to expand $E_a$ in Eq.~\eqref{eq:diffsfweightinter}
to lowest order in $W/\Delta$.
This leads to 
\begin{eqnarray}
\mathcal D^{inter}_{s}&\approx & \frac{1}{\nuc} \sum_{a \in FB}  \frac{4\Delta^2}{E_{a}}
 \Omega_I(\ket{a})\\
&= & 
\frac{1}{\nuc} \sum_{a \in FB}  \frac{4\Delta^2}{\sqrt{\Delta^2+\epsilon_a^2}}
 \Omega_I(\ket{a})\\
&= & 
\frac{1}{\nuc} \sum_{a \in FB}  \frac{4\Delta^2}{\Delta\sqrt{1+\frac{\epsilon_a^2}{\Delta^2}}}
 \Omega_I(\ket{a})\\
&= &
\frac{1}{\nuc} \sum_{a \in FB} \left[ 4\Delta - 2\frac{\epsilon_a^2}{\Delta} + \Delta \mathcal O(\frac{W^4}{\Delta^4}) \right] \Omega_I(\ket{a})\\
&\approx &\mathcal D^{inter}_{s} (W=0) + \delta \mathcal D^{inter}_{s},
\end{eqnarray}
where the disorder induced change is given by
\begin{equation}
\delta \mathcal D^{inter}_{s}= -\frac{1}{\nuc} \sum_{a \in FB}  2\frac{\epsilon_a^2}{\Delta}  \Omega_I(\ket{a}).
\end{equation}
We emphasize that this is the lowest order correction.

We now consider the two scenarios of the main text.

(i) For a single impurity, we obtain
\begin{eqnarray}
\delta \mathcal D^{inter}_{s}|_\boundstate &\approx& -\frac{1}{\nuc}  2\frac{\epsilon_\boundstate^2}{\Delta}  \Omega_I(\ket{\boundstate})\\
 &=& -\frac{2}{\nuc}  2\frac{W^2}{\Delta N_f^2}  \Omega_I(\ket{\boundstate}),
\end{eqnarray}
obtaining a result quoted in the main text. Note that in going to the last line, we performed disorder average,
using that
$\epsilon_\boundstate = \epsilon_{\mathbf r, \slind}/N_f$ and that $\disav{\epsilon_{\mathbf r, \slind}^2} = W^2$.

(ii) For the full disordered Hamiltonian, Eq.~\eqref{eq:disorderham}, we obtain 
\begin{eqnarray}
\delta \mathcal D^{inter}_{s} &\approx&  -\frac{2}{\nuc}\sum_{a \in FB} \frac{\epsilon_{a}^2}{\Delta}  \Omega_I(\ket{a})\\
&=& -\frac{2}{\nuc \Delta}\Tr\left[\hdis^2 \posop\left(1-\projfb\right)\posop   \right]\\
&=&-\frac{2W^2}{ N_f \Delta}\Omega_I(\ket{\boundstate}),
\end{eqnarray}
arising as  $\nuc N_f$ times the single impurity contribution.

\subsection{Intraband superfluid weight}
\label{suppl:intraband}
(i) For a single bound state, we calculate the intraband contribution as in the main text:
\begin{eqnarray}
\delta \mathcal D^{\text{intra}}_{s}&\approx& \frac{2}{\nuc}\sum_{s_1 \in FB,s_2 \in FB}  
\frac{1}{2\Delta}|\braket{s_1|\currop|s_2}|^2 \\
&=& \frac{1}{\nuc}\frac{1}{\Delta}  \mathrm{Tr}[\projfb \currop \projfb \currop]\\
&=&\frac{1}{\nuc}\frac{1}{\Delta}  \mathrm{Tr}\left\{\projfb [\posop,\hdis] \projfb [\hdis,\posop]\right\}\\
&=&\frac{2}{\nuc}\frac{1}{N_f^2}\frac{ W^2}{\Delta}  \tilde \Omega(\ket{\boundstate}) .
\end{eqnarray}
(ii) Considering the full disordered Hamiltonian, we use the 
independence of impurities on different sites as follows:
\begin{eqnarray}
\delta \mathcal D^{\text{intra}}_{s}&\approx& \frac{2}{\nuc}\sum_{s_1 \in FB,s_2 \in FB}  
\frac{1}{2\Delta}|\braket{s_1|\currop|s_2}|^2 \\
&=& \frac{1}{\nuc}\frac{1}{\Delta}  \mathrm{Tr}[\projfb \currop \projfb \currop]\\
&=&\frac{1}{\nuc}\frac{1}{\Delta}  \mathrm{Tr}\left\{\projfb [\posop,\projfb \sum_{\mathbf r, \slind} \epsilon_{\mathbf r,\slind} \ketbra{\mathbf r,\slind} \projfb] \projfb [\projfb \sum_{\mathbf r, \slind} \epsilon_{\mathbf r,\slind} \ketbra{\mathbf r,\slind} \projfb,\posop]\right\}\\
&=&\sum_{\mathbf r, \slind}
\frac{1}{\nuc}\frac{W^2}{\Delta}  \mathrm{Tr}\left\{\projfb [\posop,\projfb \ketbra{\mathbf r,\slind} \projfb] \projfb [\projfb  \ketbra{\mathbf r,\slind} \projfb,\posop]\right\}\\
&=&\nuc N_f
\frac{1}{\nuc}\frac{W^2}{\Delta}  \mathrm{Tr}\left\{\projfb [\posop,\projfb \ketbra{\mathbf r_{\boundstate},\slind_{\boundstate}} \projfb] \projfb [\projfb  \ketbra{\mathbf r_{\boundstate},\slind_{\boundstate}} \projfb,\posop]\right\}\\
&=&\nuc N_f  \frac{2}{\nuc}\frac{1}{N_f^2}\frac{ W^2}{\Delta}  \tilde \Omega(\ket{\boundstate}),
\end{eqnarray}
where in getting to the second to last line,
we used the assumption that impurities on different sites are symmetry related, and replaced a sum over $\alpha$
by a factor of $N_f$ times the value for a representative bound state on sublattice $\alpha_{\boundstate}$.
Within this assumption, all the bound states have the same localization and the projected
site wavefunctions have the same norm $|\projfb \ket{\mathbf r,\slind}|^2 = 1/N_f$  for flat band orbitals $\slind$.
Note that the symmetry relation also guarantees uniform pairing on the flat band orbitals.

\subsection{Equality of the intraband current for projected and unprojected disorder}
\label{suppl:projectionworks}
In the main text, we use projected disorder. Here, we derive that
the current matrix element of projected disorder and unprojected
disorder are the same to lowest order in $W/E_G$, where $E_G$ is the gap to the remote bands.
To that end, let us assume the projected problem has been solved, and flat band eigenstates labelled by index $a \in FB$ have been obtained.
The effect of the interband component of disorder is to mix these eigenstates with the remote band states.
Within first order perturbation theory, the flat band eigenstates are perturbed as
\begin{equation}
\ket{a} \to \ket{a} +  \sum_{m \neq m_{FB}, \mathbf k}\frac{1}{\epsilon_{m,\mathbf k}-\epsilon_a} \ketbra{m,\mathbf k} \hdisunproj  \ket{a},
\end{equation}
where we denote by $\hdisunproj$ the disorder Hamiltonian before flat band projection.
Using these perturbed wavefunctions, we can evaluate the intraband matrix elements, focusing on lowest order terms in $W/E_G$:
\begin{eqnarray*}
\braket{s_1|\currop|s_2}&=& 
\sum_{m \neq m_{FB}, \mathbf k}
\left[
\braket{s_1|\currop
\frac{1}{\epsilon_{m,\mathbf k}-\epsilon_{s_2}} \ketbra{m,\mathbf k} \hdisunproj  |s_2}+ 
\braket{s_1|\frac{1}{\epsilon_{m,\mathbf k}-\epsilon_{s_1}}\hdisunproj \ketbra{m,\mathbf k}
\currop|s_2}
\right]\\&=& 
\sum_{m \neq m_{FB}, \mathbf k}
i\left[
\braket{s_1|
\frac{[\mathbf r,\hclean]}{\epsilon_{m,\mathbf k}-\epsilon_{s_2}} \ketbra{m,\mathbf k} \hdisunproj  |s_2}+ 
\braket{s_1|\frac{\hdisunproj}{\epsilon_{m,\mathbf k}-\epsilon_{s_1}} \ketbra{m,\mathbf k}
[\mathbf r,\hclean]|s_2}
\right]\\&=& 
\sum_{m \neq m_{FB}, \mathbf k}
i\left[
\braket{s_1|\mathbf r
 \ketbra{m,\mathbf k} \hdisunproj  |s_2}-
\braket{s_1|\hdisunproj \ketbra{m,\mathbf k}
\mathbf r|s_2}
\right] + \mathcal{O}(W^2/E_G)
\\&=& 
i\braket{s_1|\hdisunproj (1-\projfb)\mathbf r -
\mathbf r (1-\projfb)\hdisunproj |s_2}
+ \mathcal{O}(W^2/E_G)
\\&=& 
\braket{s_1|i[\projfb \hdisunproj \projfb ,\mathbf r ]|s_2}
+ \mathcal{O}(W^2/E_G)
\\&=& 
\braket{s_1|i[\hdis ,\mathbf r ]|s_2}
+ \mathcal{O}(W^2/E_G),
\end{eqnarray*}
recovering the intraband current induced by the projected Hamiltonian.
Note that in going to the third line, we set $\epsilon_{s_i} = \epsilon_{m_{FB},\mathbf k}$, discarding higher order terms
in $W/E_G$.

\subsection{Generalization to arbitrary rank-1 disorder and analysis of Wannier disorder }
\label{suppl:generalizerankone}

Our results can be immediately extended to other types of rank-1 disorder. A general rank-1 disordered Hamiltonian
is given as 
\begin{equation}
\hdis^{\text{rank 1}} =  \sum_{j} \epsilon_{j} \frac{1}{N_j} \ket{j}\bra{j} ,
    \end{equation}
where the $\ket{j}$ are arbitrary wavefunctions within the flat band subspace and where the different $\epsilon_j$ are uncorrelated, having variance $\disav{\epsilon_j^2}= W^2$, and where $N_j$ are arbitrary coefficients.
Due to the uncorrelated nature of the disorder, the superfluid weight change is a sum due to each individual term
\begin{equation}
\delta \mathcal D_{s} = 2\frac{W^2 }{\nuc  \Delta}\sum_j \frac{1}{N_j^2} \left[\tilde \Omega(\ket{j})-\Omega_I(\ket{j})\right],
\end{equation}
using the same arguments as above for on-site disorder.

The most physically relevant example of such rank-1 disorder is on-site disorder that was the focus of the main text. 
Another example is Wannier disorder, having the Hamiltonian
\begin{equation}
\hdis =  \sum_{\mathbf r} \epsilon_{\mathbf r} \ket{m_{FB},\mathbf r}\bra{m_{FB},\mathbf r} ,
    \end{equation}
where $\ket{m_{FB},\mathbf r}$ is a Wannier function of the flat band centered on site $\mathbf r$,
and where the Wannier function energies $\epsilon_{\mathbf r}$ are independent, identically-distributed random variables of zero mean
and variance $\disav{\epsilon_{\mathbf r}^2 } = W^2$.
The change in superfluid weight is related to the two terms in the 
localization functional for the Wannier function, given as
\begin{equation}
\delta \mathcal D^{\text{Wannier}}_{s} = 2\frac{W^2}{ \Delta}\left[\tilde \Omega(\ket{m_{FB},\mathbf r})-\Omega_I(\ket{m_{FB},\mathbf r})\right],
\end{equation}
a competition between the gauge-invariant and gauge-dependent parts of the localization functional for the Wannier function that defines the disorder.

The relative magnitude of the two parts of the functional depends strongly on the system considered.
In one dimension, $\tilde \Omega(\ket{m_{FB},\mathbf r})=0$ for maximally localized Wannier functions \cite{vanderbiltMarzariMaximallyLocalizedWannier2012}.
In two-dimensional bands with nonzero Berry curvature, 
$\tilde \Omega$ is typically nonzero even for maximally localized Wannier functions.
For instance, in a topological band,  $\tilde \Omega(\ket{m_{FB},\mathbf r}) = \infty $ as Wannier functions 
in such bands cannot be exponentially localized. However, while this naively predicts a large increase in superfluid 
weight due to disorder, we consider such long-range disorder to be unphysical.
Note that there are no problems in connecting $\tilde{\Omega}$ with an observable quantity like superfluid weight, although for Wannier functions it is not gauge-invariant: here, a particular choice of Wannier functions simply means a particular choice of disorder.

\section{Numerical methods}
\subsection{Details of the tight-binding models}
\label{suppl:tbdetails}
\begin{figure}[t]
    \includegraphics[width=0.5\columnwidth]{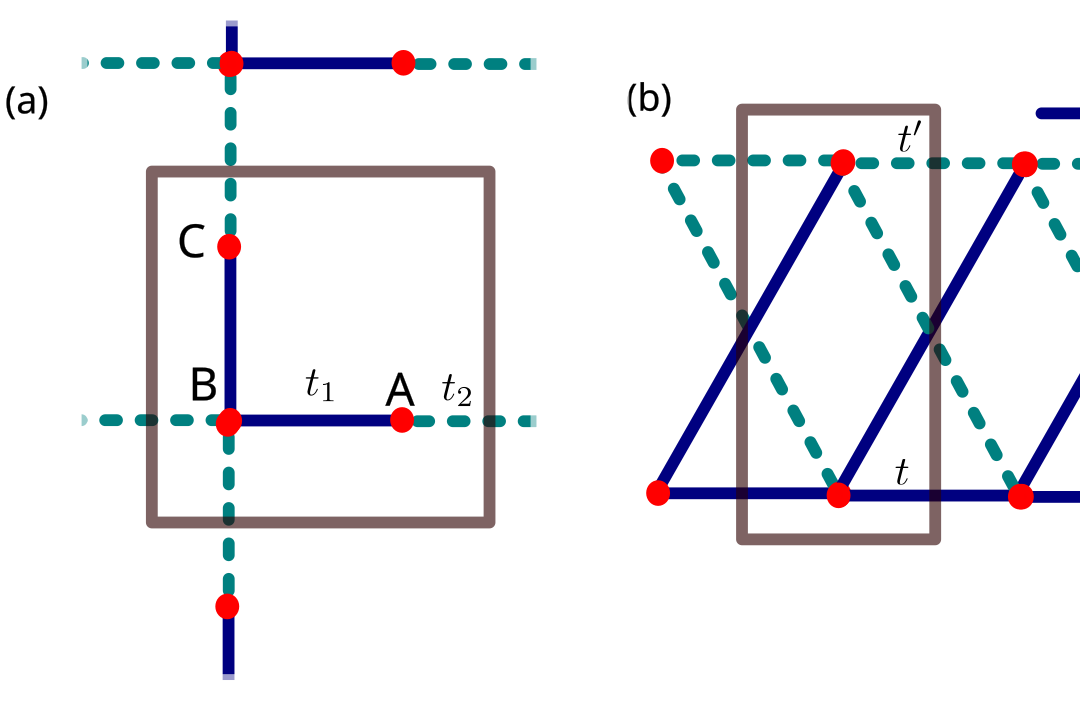}
    \caption{ Schematic of the Lieb (a) and Creutz (b) lattices used. The unit cell is enclosed in a grey frame. 
    }
    \label{fig:figlattices}
\end{figure}

For the three-orbital two-dimensional Lieb lattice, shown in Fig.~\ref{fig:figlattices}a, there are
three bands, and the middle one, denoted by $m_{FB}=2$, is flat. 
We work at staggered hoppings $t_1=1$ and $t_2=0.8$ and chemical potential $\mu=0$, so that the flat band is at
zero energy, having $\epsilon_{2,\mathbf k}=0$.
We parametrize the orbital positions for the Lieb lattice as
$\mathbf r_\alpha = x \mathbf r^{(0)}_\alpha,$
where $\mathbf r^{(0)}_\alpha$ are the positions shown in Fig.~\ref{fig:figlattices}a, where the origin of the unit cell is taken to be at the $B$ site. Explicitly, they are given as 
$\mathbf r^{(0)}_B = [0,0]$, $\mathbf r^{(0)}_A =[0.5,0]$ and $\mathbf r^{(0)}_C= [0,0.5]$.
For $x=0$, all the orbitals are on top of each other, while for $x=1$ we recover the positions $\mathbf r^{(0)}_\alpha$.
For the Lieb lattice at these parameters, the minimal quantum metric is obtained for $x=0.197$.

For the Creutz ladder, shown in Fig.~\ref{fig:figlattices}b, the hoppings have the values $t= 1$ and $t'=-1$ and there
are two flat bands. Setting the chemical potential $\mu=-2$, the 
bottom flat band is at zero energy $\epsilon_{1,\mathbf k}=0$,
while the upper flat band has $\epsilon_{2,\mathbf k}=4$. The quantum metric is uniform for this model.

For the above tight-binding models, we write $\hclean$ in real-space as 
\begin{equation}
\hclean  = \sum_{\mathbf r,\alpha,\mathbf r',\beta}  \ket{\mathbf r,\alpha}t_{\mathbf r\alpha,\mathbf r'\beta} \bra{\mathbf r',\beta}.
\end{equation}
When a vector potential $\vecpot$ is applied, the hoppings are changed according to Peierls substitution as
\begin{equation}
\hclean(\vecpot)  = \sum_{\mathbf r,\alpha,\mathbf r',\beta}  \ket{\mathbf r,\alpha}e^{i\vecpot \cdot (\mathbf r+\mathbf r_\alpha-\mathbf r'-\mathbf r_\beta)}
t_{\mathbf r\alpha,\mathbf r'\beta} \bra{\mathbf r',\beta},
\end{equation}
where $\mathbf r_\alpha$ is the position of orbital $\alpha$ within the unit cell.

\subsection{Calculation of localization functionals}

\label{suppl:locfuncdetails}

We calculate the localization functionals listed in Table~\ref{tab:table1} of the main text
in real-space.
For the Creutz ladder, both of them can be analytically shown to be equal to $1/4$.
To numerically evaluate $\tilde\Omega(\ket{\boundstate})$ and $\Omega_I(\ket{\boundstate})$ for the Lieb lattice, we add a single on-site impurity in the center of a finite sample, and calculate the intra and interband spread of the bound state wavefunction following Eq.~\eqref{eq:locfunctionalI} and Eq.~\eqref{eq:locfunctionalII}.
We find that the bound state wavefunctions are well-localized and that the results quickly converge upon increasing system size, saturating at $\nuc=400$ unit cells. 
We calculate the localization functionals for the orbital positions that correspond to the minimal quantum metric~\cite{tormaHuhtinenRevisitingFlatBand2022}, as these are the relevant ones for the superfluid weight calculation. 

We observe an interesting orbital position dependence of $\tilde\Omega(\ket{\boundstate})$ and $\Omega_I(\ket{\boundstate})$ for the Lieb lattice, shown in Fig.~\ref{fig:supplorbitaldependence}. 
Intriguingly, the minima of $\tilde \Omega$ and $\Omega_I$ coincide, and at the minimum they are equal.
Note that because for the Lieb lattice the two orbitals on which the flat band is supported are related by symmetry, 
the bound state interband localization is equal to the Wannier function localization: $\Omega_I(\ket{\boundstate}) =  \Omega_I(\ket{WF})$, which is simply the integral of the quantum metric. 
This fact is shown as follows. Consider the $A$ and $C$ sublattice bound states on the flat band.
They are mapped onto each other by symmetry, therefore their localization functionals must be equal.
To show that the bound state localization is also equal to the integrated quantum metric, we use the decomposition
of Eq.~\eqref{eq:eigenstatesuperposition} for the unnormalized flat band projected orbitals:
\begin{eqnarray}
\projfb\ket{\mathbf r=0,A} = \int d \mathbf k \frac{f_A(\mathbf k)}{(2\pi)^d}  \ket{\Psi(m_{FB},\mathbf k)} \\
\projfb\ket{\mathbf r=0,C} = \int d \mathbf k \frac{f_C(\mathbf k)}{(2\pi)^d}  \ket{\Psi(m_{FB},\mathbf k)} .
\end{eqnarray}
Importantly, the flat band is spanned by only the $A$ and $C$ orbitals -- it has zero weight on the $B$ orbital.
The coefficients of the above expansion are given as $f_A(\mathbf k) = \braket{\Psi(m_{FB},\mathbf k)|\mathbf r=0,A}$ and 
$f_C(\mathbf k) = \braket{\Psi(m_{FB},\mathbf k)|\mathbf r=0,C}$,
which implies that 
\begin{equation}
|f_A(\mathbf k)|^2 + |f_C(\mathbf k)|^2 = 1,
\end{equation}
using the fact that the flat band wavefunctions are nonzero only on the $A$ and $C$ orbitals.
Next, the symmetry relating $A$ to $C$ implies that 
\begin{equation}
\int d \mathbf k \frac{|f_A(\mathbf k)|^2}{(2\pi)^d}= 
\int d \mathbf k \frac{|f_C(\mathbf k)|^2}{(2\pi)^d}  =\frac{1}{N_f} = \frac{1}{2},
\end{equation}
reflecting the unnormalized nature of the states. To normalize, we need to multiply the projected orbitals by $N_f$,
defining the normalized kets as
$\ket{\boundstate_A} =\sqrt{N_f}\projfb\ket{\mathbf r=0,A}$, 
$\ket{\boundstate_C} =\sqrt{N_f}\projfb\ket{\mathbf r=0,C}$. 
The sum of the corresponding localization functionals is equal to 
\begin{eqnarray}
\Omega_I(\ket{\boundstate_A}) +  \Omega_I(\ket{\boundstate_C})  &=&
N_f\int d \mathbf k \frac{|f_C(\mathbf k)|^2+|f_A(\mathbf k)|^2}{(2\pi)^d} \mathrm{Tr}\, g(\mathbf k)   \\
&=&
N_f\int d \mathbf k \frac{1}{(2\pi)^d} \mathrm{Tr}\, g(\mathbf k)   \\
&=&
2\Omega_I(\ket{WF}).
\end{eqnarray}
However, since the localization functionals for the two bound states are equal by symmetry, it necessarily follows
that they are each equal to the Wannier localization (integrated quantum metric): 
$\Omega_I(\ket{\boundstate_A}) = \Omega_I(\ket{\boundstate_C}) = \Omega_I(\ket{WF})$.

\begin{figure}[t]
    \includegraphics[width=0.5\columnwidth]{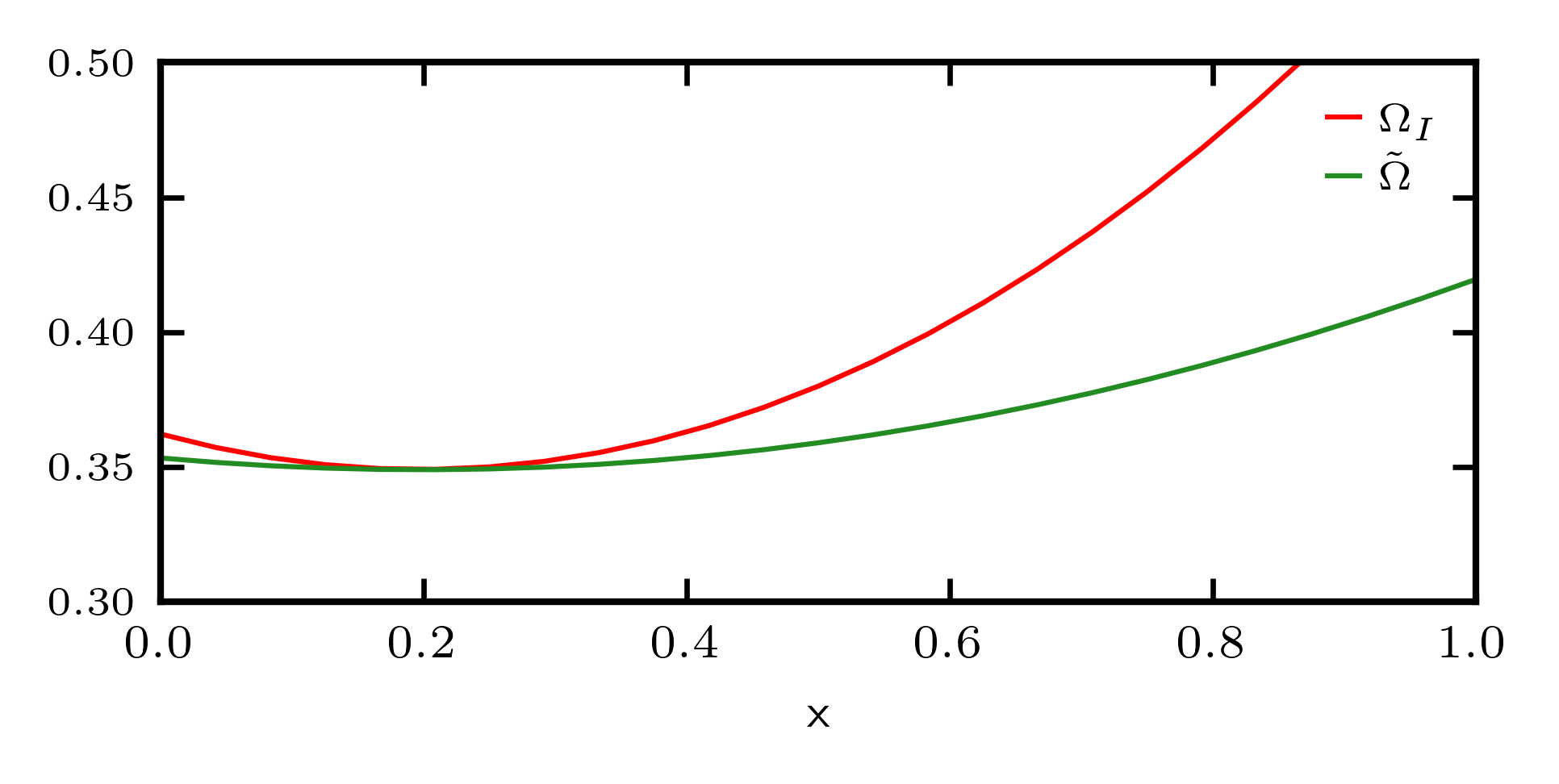}
    \caption{ The localization functionals $\Omega_I(\ket{\boundstate})$ and $\tilde \Omega(\ket{\boundstate})$ as a function of orbital positions
    parametrized by $x$ in the unit cell. Here $x=0$ corresponds to all orbitals being at the same point, while $x=1$ corresponds
    to the orbitals arranged as in Fig.~\ref{fig:figlattices}a.
    }
    \label{fig:supplorbitaldependence}
\end{figure}
\subsection{Details on the full self-consistent disordered calculation}
\label{suppl:numericsdetails}
We numerically evaluate the disordered superfluid weight within a self-consistent 
mean-field calculation with an attractive Hubbard interaction of strength $U=-1$ (in the units of the hopping strength) acting as the pairing glue, working at zero temperature.
First, we write the kinetic and disorder Hamiltonian (referred to as single-particle (SP) Hamiltonian) for the two spin species in many-body language as
\begin{equation}
\hsp =  \Psi^\dagger_{\uparrow}\htot \Psi^{\phantom \dagger}_{\uparrow}
+  \Psi^\dagger_{\downarrow}\trs \htot \trs^{-1} \Psi^{\phantom \dagger}_{\downarrow}
\end{equation}
where $\Psi^\dagger_s|_{\mathbf r,\alpha} = c^\dagger_{\mathbf r,\alpha,s}$ is a vector of creation operators for spin $s$
for each position and sublattice, and where $\trs$ is the time-reversal operator, acting in position representation as complex conjugation.
The Hamiltonian $\htot=\hclean+\hdis$ is the sum of the clean and disordered parts.

The interaction reads:
\begin{equation}
\label{eq:hamint}
\hint= U \sum_{\mathbf r, \slind}  c^\dagger_{\mathbf r, \slind,\uparrow}c^\dagger_{\mathbf r, \slind,\downarrow} 
c^{\phantom{\dagger}}_{\mathbf r, \slind,\downarrow}c^{\phantom{\dagger}}_{\mathbf r, \slind,\uparrow}.
\end{equation}

To calculate the superfluid weight, which is defined as the derivative of the grand potential with respect
to an externally applied vector potential, we add a small uniform vector potential (gauge field) $\vecpot$ to the single-particle Hamiltonian via Peierls substitution, 
\begin{equation}
\hsp(\vecpot) =  \Psi^\dagger_{\uparrow}\htot(\vecpot) \Psi^{\phantom \dagger}_{\uparrow}
+  \Psi^\dagger_{\downarrow}[\trs \htot\trs^{-1}](\vecpot) \Psi^{\phantom \dagger}_{\downarrow},
\end{equation}
where the vector potential $\vecpot$ enters $\htot(\vecpot)$ via Peierls substitution as
\begin{equation}
\htot(\vecpot)  = \sum_{\mathbf r,\alpha,\mathbf r',\beta}  \ket{\mathbf r,\alpha}e^{i\vecpot \cdot (\mathbf r+\mathbf r_\alpha-\mathbf r'-\mathbf r_\beta)}
\braket{\mathbf r,\alpha| \htot|\mathbf r',\beta} \bra{\mathbf r',\beta}.
\end{equation}
When expanded to lowest two orders in $\vecpot$,
this leads to the paramagnetic and diamagnetic terms
\begin{equation}
\label{hgetscurrentop}
 \htot(\vecpot) =\htot+ \currop \cdot \vecpot +  \frac{1}{2} \vecpot \cdot \hat T_{dia} \cdot \vecpot +  \mathcal{O} (\vecpot^3),
\end{equation}
where the paramagnetic operator is $\currop = i [\posop,\htot]$,
and the diamagnetic one is $\hat T_{dia} = -i [\posop,\currop]$, which is to be understood as a tensor in the real space coordinate index.

We treat the interaction term in the mean-field approximation, decoupling it on every site as
$  Uc^\dagger_{\mathbf r, \slind,\uparrow}c^\dagger_{\mathbf r, \slind,\downarrow} 
c^{\phantom{\dagger}}_{\mathbf r, \slind,\downarrow}c^{\phantom{\dagger}}_{\mathbf r, \slind,\uparrow} \to 
\left[ c^\dagger_{\mathbf r, \slind,\uparrow}c^\dagger_{\mathbf r, \slind,\downarrow} 
\Delta_{\mathbf r,\slind} +\hc\right] -\frac{|\Delta_{\mathbf r,\slind}|^2}{U}$,
where the mean-field order parameter is given as
\begin{equation}
\Delta_{\mathbf r,\slind} = U\mfav{ c^{\phantom{\dagger}}_{\mathbf r, \slind,\downarrow}c^{\phantom{\dagger}}_{\mathbf r, \slind,\uparrow} } ,
\end{equation}
being diagonal in position and sublattice because of the on-site nature of the Hubbard attraction.
Note that $\mfav{\hat O}$ denotes the expectation value of operator $\hat O$ in the mean field state.
The mean-field Hamiltonian is 
\begin{equation}
\label{eq:hmfsuppl}
\hmf =  \hsp(\vecpot)+
\sum_{\mathbf r,\slind}\left\{\left[ c^\dagger_{\mathbf r, \slind,\uparrow}c^\dagger_{\mathbf r, \slind,\downarrow} 
\Delta_{\mathbf r,\slind} +\hc\right] -\frac{|\Delta_{\mathbf r,\slind}|^2}{U}\right\},
\end{equation}
and we are looking for self-consistent ground states of this Hamiltonian.
Starting with an initial ansatz, we calculate the mean-field Hamiltonian [Eq.~\eqref{eq:hmfsuppl}] and
fill all the negative energy states, forming a new trial state, which is then used to calculate the mean-field Hamiltonian 
for the next iteration.
To optimize convergence, we use the optimal damping algorithm \cite{lebrisCancesCanWeOutperform2000}.
At self-consistency, the converged state is a mean-field state that minimizes the grand potential
\begin{equation}
\Omega^{\text{Grand}}(\mathbf A) = \mfav{ \hmf}.
\end{equation}
Note that in this approach, we are solving for the electronic state.
An equivalent approach is to minimize the grand potential with respect to the order parameters 
$\Delta_{\mathbf r,\slind}$ \cite{tormaHuhtinenRevisitingFlatBand2022}.

Using the grand potential, the superfluid weight is given as
\begin{equation}
\mathcal D_{s,ij} =\frac{1}{\nuc} \frac{d \Omega^{\text{Grand}}(\mathbf A)}{d A_i dA_j} .
\end{equation}
Taking a total derivative instead of partial one is crucial, as only then is the superfluid weight
orbital-position independent~\cite{tormaHuhtinenRevisitingFlatBand2022}.

We work in a finite periodic system. 
For the Creutz ladder we use $\nuc=200$ unit cells and average over $20$ disorder realizations for the main text data.
 See Extended data Section~\ref{suppl:extendeddata} for system size dependence of various quantities for the Creutz ladder. 
For the Lieb lattice, we study a square system totalling $\nuc=196$ unit cells, and average over $20$ different random realizations of disorder for the main text data.
\subsubsection{Hartree term}
The attractive Hubbard interaction 
Eq.~\eqref{eq:hamint}
can also be decoupled in the
Hartree channel, leading to self-consistent local electrostatic potentials.
The mean-field Hartree Hamiltonian then reads
\begin{equation}
\hint^{Hartree}= U \sum_{\mathbf r, \slind}  c^\dagger_{\mathbf r, \slind,\uparrow}c^{\phantom{\dagger}}_{\mathbf r, \slind,\uparrow}
\langle c^\dagger_{\mathbf r, \slind,\downarrow}  c^{\phantom{\dagger}}_{\mathbf r, \slind,\downarrow} \rangle  + 
c^\dagger_{\mathbf r, \slind,\downarrow}c^{\phantom{\dagger}}_{\mathbf r, \slind,\downarrow}
\langle c^\dagger_{\mathbf r, \slind,\uparrow}  c^{\phantom{\dagger}}_{\mathbf r, \slind,\uparrow} \rangle.
\end{equation}
This term affects the result in the main text, primarily through promoting variations
in local density. These act to reduce $\ds$.
\hartreediscussion{}

\begin{figure}[t]
    \includegraphics[width=0.9\columnwidth]{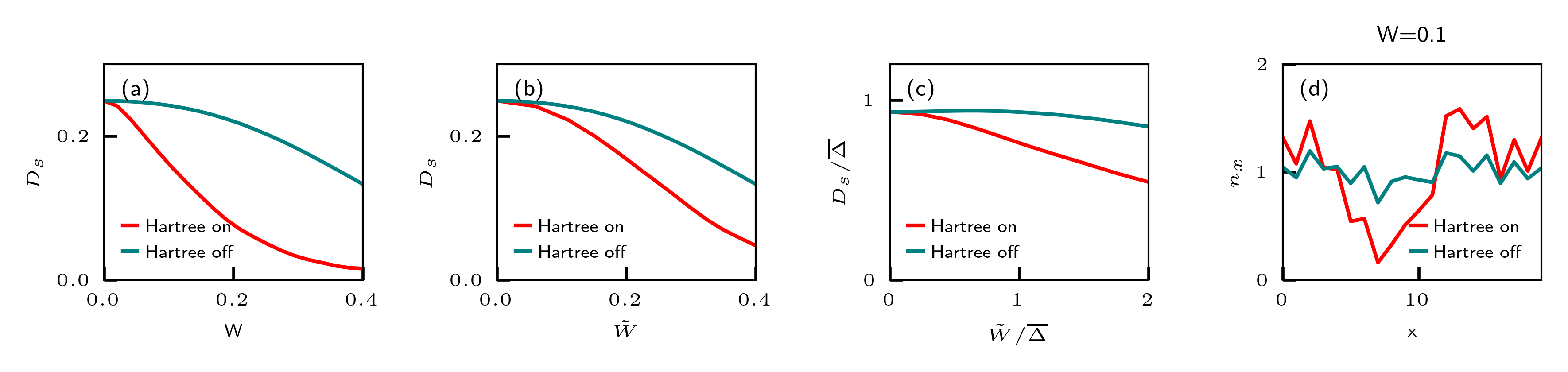}
    \caption{\creutzhartreecap}
    \label{fig:effectofhartree}
\end{figure}

\begin{figure}[t]
    \includegraphics[width=0.9\columnwidth]{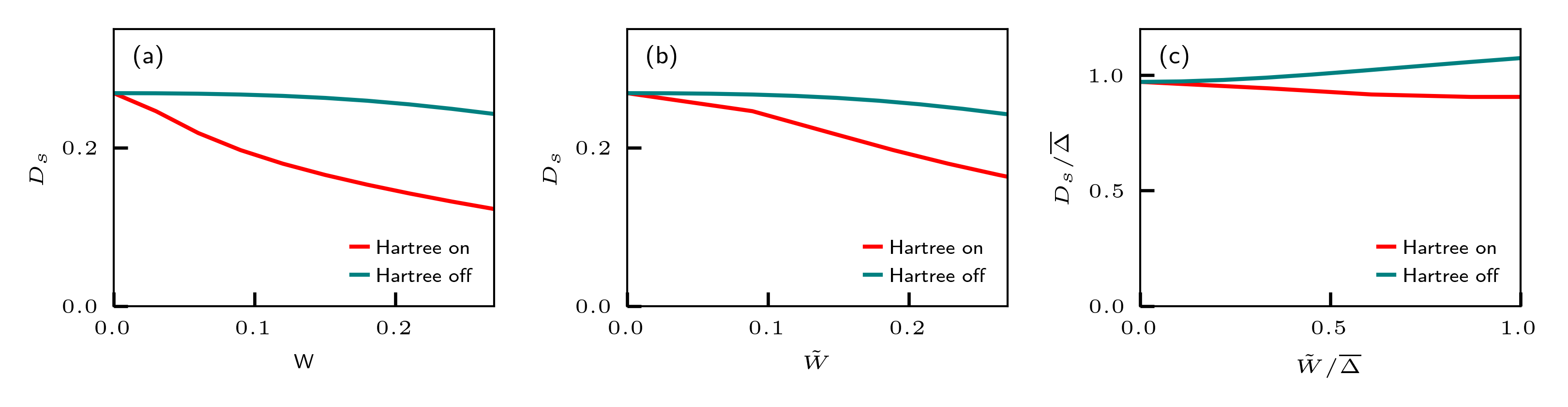}
    \caption{\liebhartreecap }
    \label{fig:effectofhartreelieb}
\end{figure}

\subsubsection{Role of the filling}
Our results were derived in the grand-canonical ensemble where the chemical
potential is adjusted so that the flat band is at zero energy,
which corresponds to half-filling of the flat band.
Note that the total filling of the system is the sum of the 
flat band filling and the filling of all the other bands in the system.
Half-filling of the flat band has the advantage that when disorder is added, 
the half-filling is preserved on average.
This holds exactly for systems with particle-hole symmetry around the flat band (such as the Lieb lattice)
when disorder symmetric around $E=0$ is added, and approximately for
models without particle-hole symmetry around the flat band (such as the Creutz ladder), provided we are in the isolated band limit.
Importantly, this is not the case for other fillings
of the flat band, where the flat band is at finite energy.
Furthermore, for fillings other than half-filling of the flat band, the chemical potential needs to be adjusted when disorder is added to keep the filling constant.
While this adjustment can be incorporated perturbatively close to half-filling, at general fillings, this is not possible.
Thus, we focus on the half filling of the flat band case as it does not involve these additional complications. 
We note furthermore that the mean-field flat band superfluid weight is maximal in the half-filled attractive Hubbard model \cite{tormaPeottaSuperfluidityTopologicallyNontrivial2015}, motivating the focus on half-filling.

\subsection{Isolating the intraband contribution within self-consistent numerics}
\label{suppl:removeintradetails}
To remove the intraband contribution to the superfluid weight within the self-consistent numerics, we need to modify the current operator and remove its intraband components.

In Eq.~\eqref{hgetscurrentop} we need to replace for the paramagnetic term,
$\currop \to (\currop -\projfb \currop \projfb)$,
while for the diamagnetic term, we need to replace
$\hat T_{dia} \to \hat T_{dia} + i[\projfb \posop \projfb,\projfb \currop \projfb]$.
Thus, the total Hamiltonian for the up spin without the intraband term is
\begin{equation}
 \htot^{\text{No intraband}}(\vecpot) =\htot(\vecpot)- \projfb\currop \projfb\cdot \vecpot +  \frac{i}{2} \vecpot \cdot [\projfb \posop \projfb,\projfb\currop\projfb] \cdot \vecpot.
\end{equation}
We note that for the finite-size periodic systems that we consider, there is a subtlety, as
$\projfb \posop \projfb$ itself requires a definition of origin, and jumps at the system boundaries.
We obtain a well-defined answer by
separating the current operator into a sum of different terms for each impurity,
and defining the position operator with respect to that impurity's position.
This prescription ensures
that the boundaries of the finite-size system do not affect the result.
\section{Extended data}
\label{suppl:extendeddata}
\subsection{System size dependence}
\begin{figure}[t]
    \includegraphics[width=0.95\columnwidth]{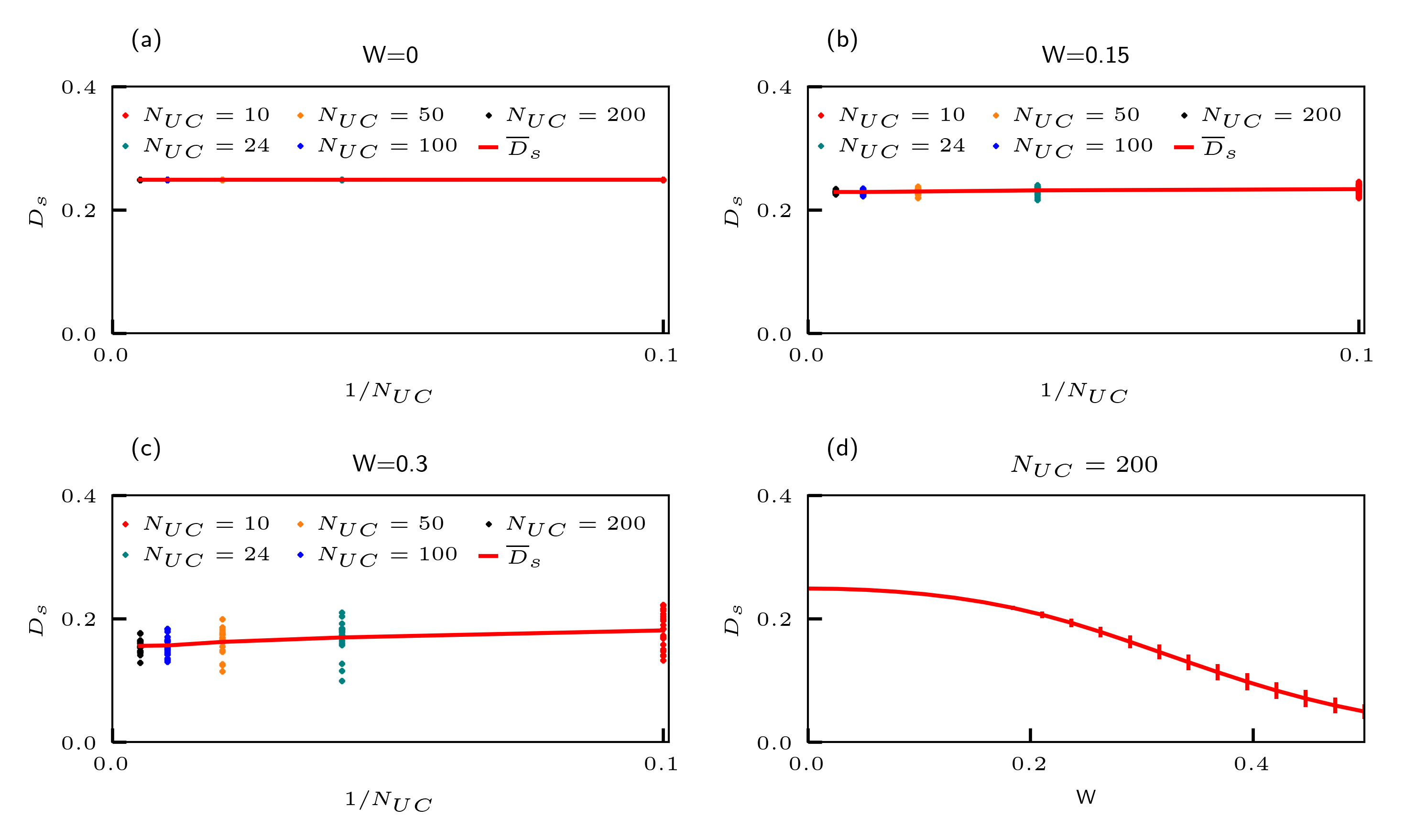}
    \caption{\dsstatscap}
    \label{fig:dsstatistics}
\end{figure}

For the system-size dependence for the Creutz ladder, see Fig.~\ref{fig:dsstatistics}.
\subsection{Parameter dependence for Lieb lattice}
For the parameter dependence for the Lieb lattice, see Figs.~\ref{fig:liebintdep}
and~\ref{fig:liebparamdep}.
\begin{figure}[t]
    \includegraphics[width=0.95\columnwidth]{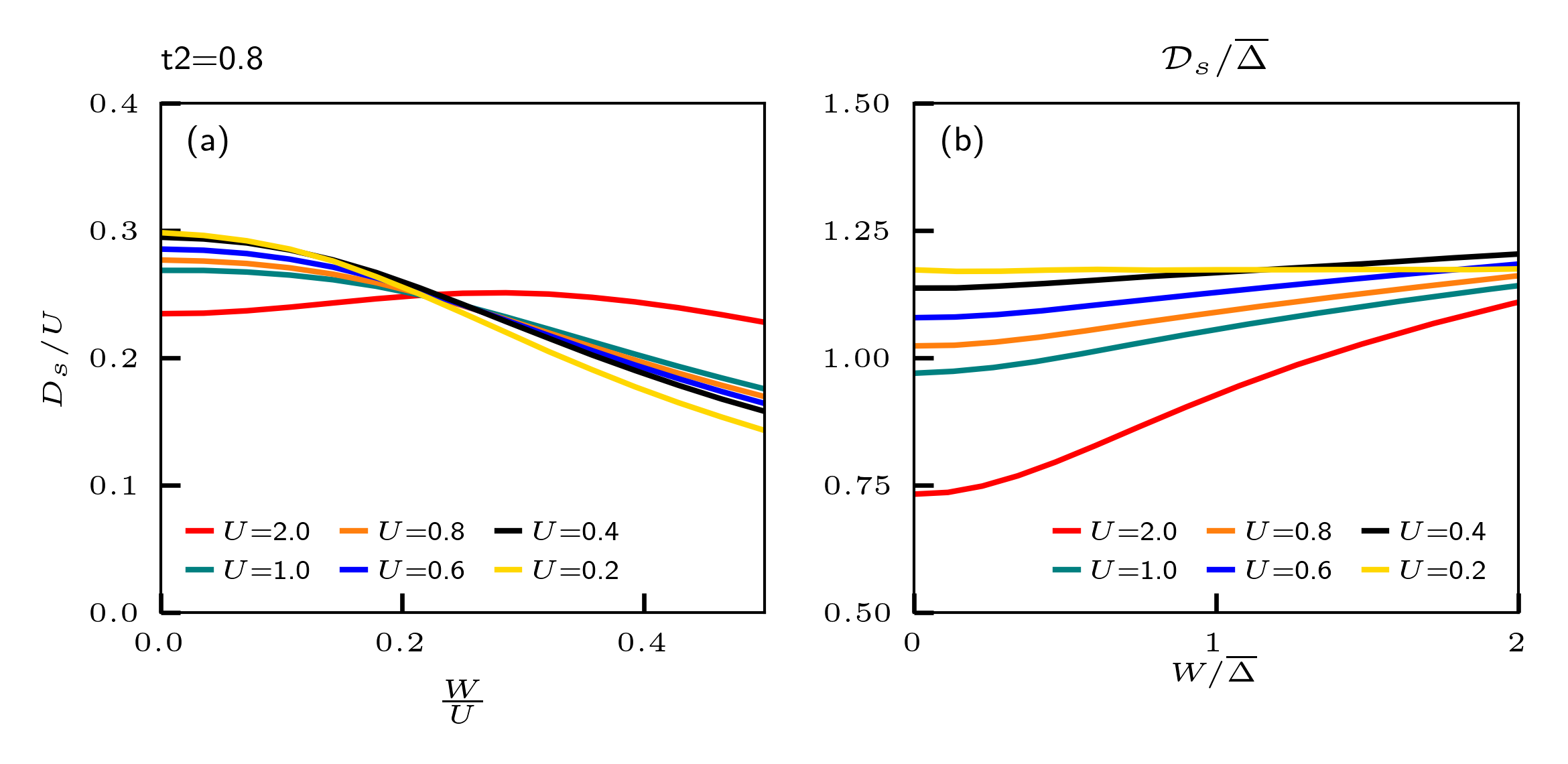}
    \caption{\liebintcap}
    \label{fig:liebintdep}
\end{figure}
\begin{figure}[t]
    \includegraphics[width=0.95\columnwidth]{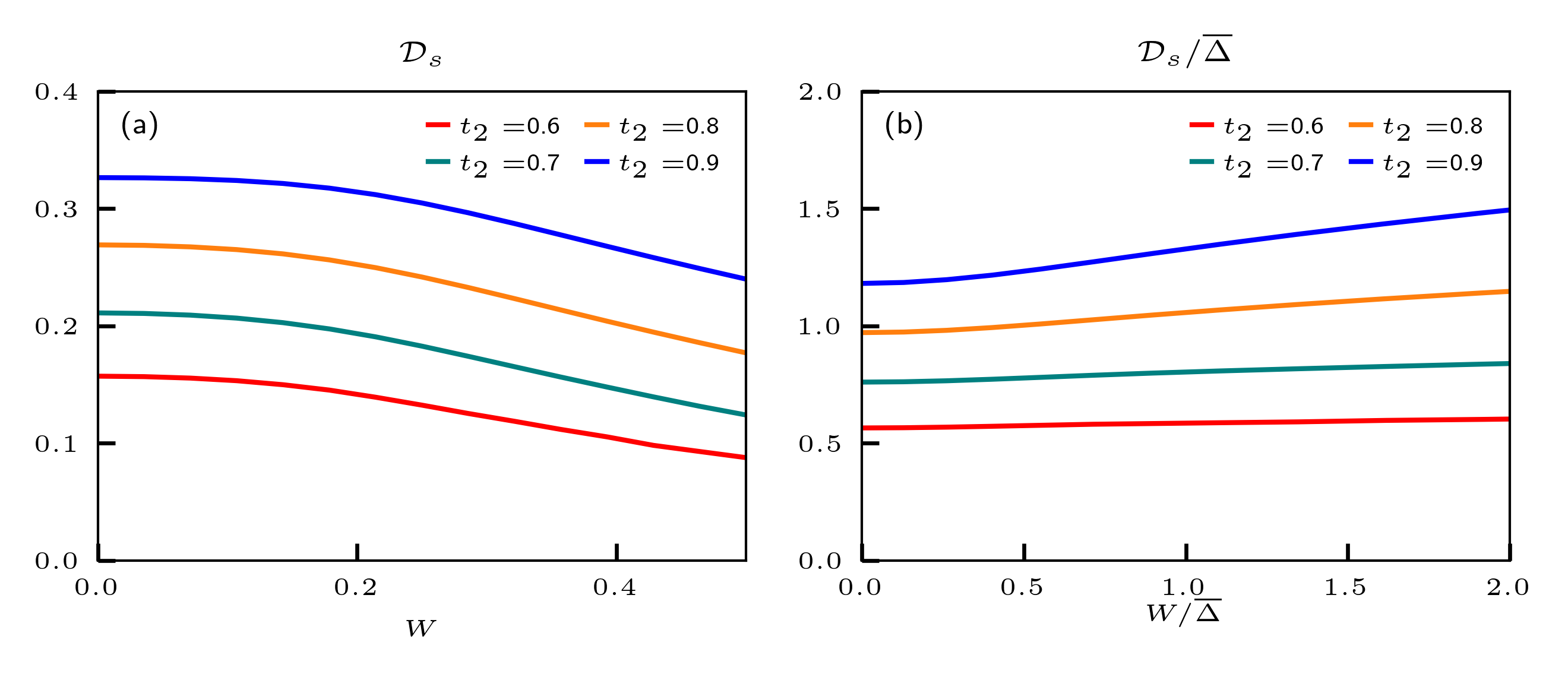}
    \caption{\liebparamcap}
    \label{fig:liebparamdep}
\end{figure}
\clearpage
\subsection{Example of a single disorder realization}
Fig.~\ref{fig:potprofile} shows the potential profile for the Creutz ladder for a single-disorder realization.
\begin{figure}[t]
    \includegraphics[width=0.5\columnwidth]{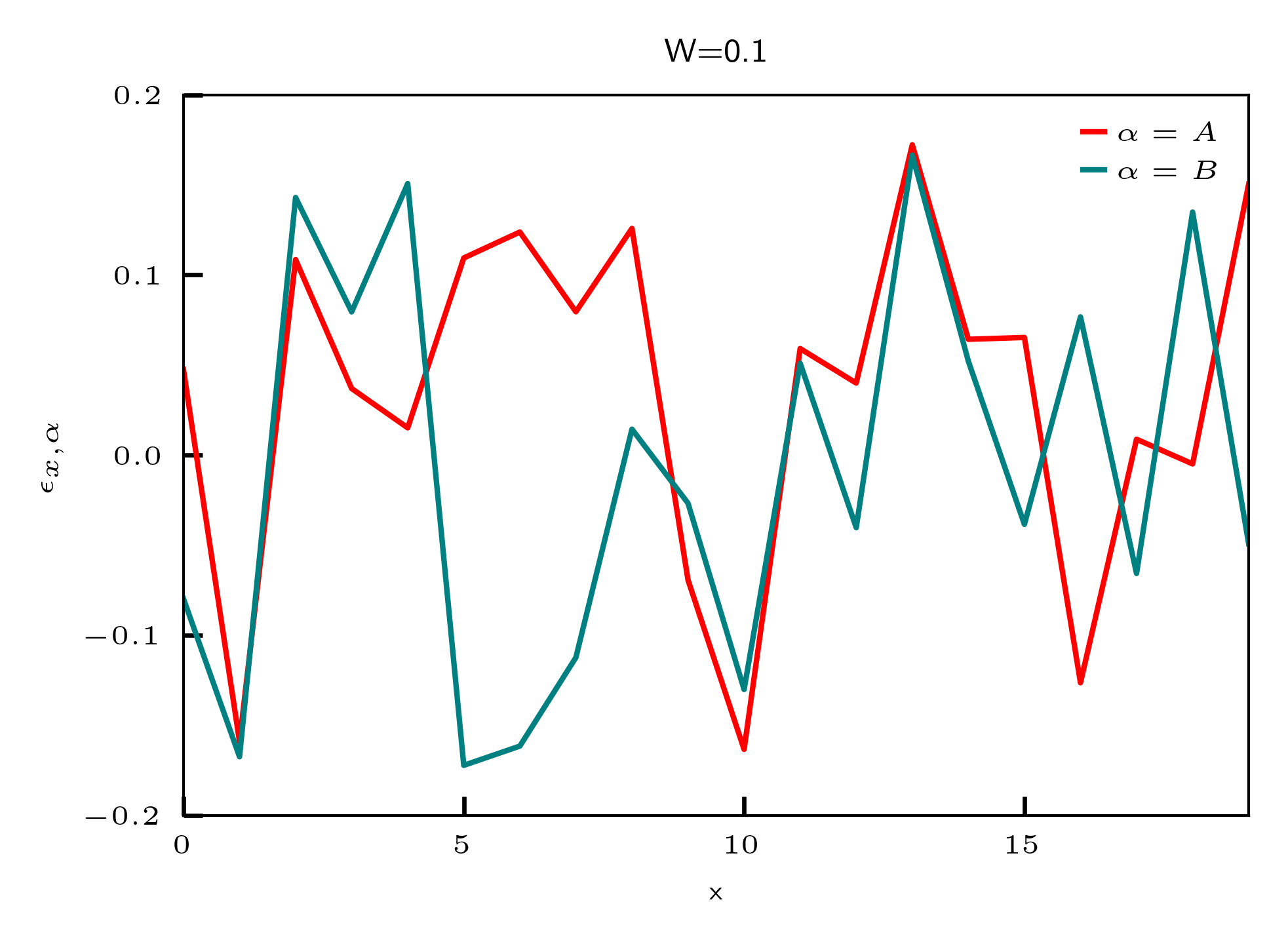}
    \caption{Potential profile for a single disorder realization for a 
    $\nuc=20$ site Creutz ladder at $W=0.1$.}
    \label{fig:potprofile}
\end{figure}
\end{widetext}

\end{document}

%% file: kmacros.tex
\def\gapmat{\hat \Delta}
\def\gap{{\Delta}}
\def\hkin{H(\mathbf k)}

\def\hbdg{H_{BdG}}
\def\Tr{\mathrm{Tr}}
\def\hdis{\hat H_{\text{dis}}}
\def\hdisunproj{\hat H^{\text{Unprojected}}_{\text{dis}}}
\def\hclean{\hat H_{\text{clean}}}
\def\hint{\hat H_{\text{int.}}}
\def\hsp{\hat H_{\text{sp}}}
\def\hmf{\hat H_{\text{MF}}}
\def\ds{\mathcal D_{s}}
\def\htot{\hat H}
\def\slind{\alpha}

\def\disindex{{s}}

\def\posop{\hat{\mathbf{r}}}
\def\currop{\hat{\mathbf{j}}}
\def\vecpot{{\mathbf{A}}}

\def\boundstate{\mathrm{BS}}
\newcommand{\disav}[1]{\left\langle #1 \right\rangle_{\text{dis}}}
\newcommand{\mfav}[1]{\left\langle #1 \right\rangle_{\text{MF}}}
\newcommand{\ketbra}[1]{\ket{#1}\bra{#1} }

\newcommand{\trs}{\mathcal{T}}

\newcommand{\colora}{red}
\newcommand{\colorb}{teal}

\newcommand{\removeKK}[1]{}

\newcommand{\nuc}{N_{\text{UC}}}
\newcommand{\norb}{N_{\slind}}

\newcommand{\hc}{h.\,c.}

\newcommand{\TODO}[1]{} 
\newcommand{\projfb}{\hat P_{FB}}

\newcommand{\creutzhartreecap}{
    Effect of the Hartree term from interactions for the Creutz ladder. 
Note that here we average over $5$ disorder realizations at $\nuc=20$.
    (a) Superfluid weight as a function of bare disorder strength $W$. The value with Hartree on (red) is notably
    reduced compared to the value without Hartree (teal).
    (b) Showing the same as (a) but as a function of renormalized disorder $\tilde W$
defined in Eq.~\eqref{eq:defrenormdisorder}.
    As a function of renormalized disorder, the decrease is not as strong as (a).
    (c) Superfluid weight divided by the order parameter as a function of renormalized disorder $\tilde W/\overline \Delta$
    defined in Eq.~\eqref{eq:defrenormdisorder}. 
    Note that even with Hartree, the superfluid weight divided by the order parameter is robust up to $\tilde W/\overline \Delta \approx 0.5$, in agreement with our analytical theory. 
    (d) Local density $n_x$ as a function of position with and without Hartree at $W=0.1$. 
    When Hartree is included, 
    charges are nearly depleted at certain positions, 
    forming weak links and suppressing the superfluid
    response of the whole system.
}

\newcommand{\liebhartreecap}{
    Effect of the Hartree term from interactions for the Lieb lattice. 
    (a) Superfluid weight as a function of bare disorder strength $W$. The value with Hartree on (red) is notably
    reduced compared to the value without Hartree (teal).
    (b) Showing the same as (a) but as a function of renormalized disorder $\tilde W$
defined in Eq.~\eqref{eq:defrenormdisorder}.
    As a function of renormalized disorder, the decrease is not as strong as (a).
    (c) Superfluid weight divided by the order parameter as a function of renormalized disorder $\tilde W/\overline \Delta$
    defined in Eq.~\eqref{eq:defrenormdisorder}. 
    Note that even with Hartree, the superfluid weight divided by the order parameter is extremely robust up to $\tilde W/\overline \Delta \approx 1$, extending beyond the strict regime of validity of our perturbative analytical theory. 
Note that here we consider only one disorder realization at $\nuc=196$.

    }

\newcommand{\dsstatscap}{
(a)  System-size dependence of $\ds$ for the Creutz ladder at $W=0$.
    Each disorder realization (in total
20 are taken for each value of $\nuc$) is shown with a scatter plot. 
For $W=0$, there is no system size dependence in the superfluid weight.
(b) Same as (a) but for $W=0.15$.
(c) Same as (a) but for $W=0.3$.
For increasing $\nuc$, the relative variation between different disorder realizations decreases, and the mean value is nearly constant.
(d) Superfluid weight $\ds$ at $\nuc=200$ as a function of $W$ with
error bars obtained from $20$ disorder realizations.
The magnitude of the error increases with $W$, but remains small.}

\newcommand{\liebparamcap}{Dependence on the hopping strength $t_2$ for the Lieb lattice at fixed interaction strength $U=1$ and with $t_1=1$ throughout. (a) The superfluid weight $\ds$ as a function of disorder strength $W$.
    For larger $t_2$, the quantum metric of the Lieb lattice flat band is increased, and the superfluid weight 
    follows this trend.
    (b) The superfluid weight divided by the average order parameter $\overline \Delta$.
Note that here we consider only one disorder realization at $\nuc=14^2$.
    }
\newcommand{\liebintcap}{Interaction strength dependence for the Lieb lattice at $t_2=0.8$. (a) Superfluid weight as a function disorder strength. Both are written in units of $U$. 
    In the $U\to 0$ limit, the curves are expected to be independent of $U$. 
    (b) The superfluid weight divided by the average order parameter $\overline \Delta$. 
Here we average over $5$ disorder realizations at $\nuc=196$.}

\newcommand{\hartreediscussion}{
Physically, the Hartree term increases the inhomogeneity created by the disorder potential, enhancing the effective value of $W$. 
To check this intuition, we have compared simulations with and without this term.
First, we consider the 1D Creutz ladder. 
In one dimension, the Hartree term could be particularly detrimental, as any zero-density region that might arise due to it could effectively block any supercurrent from passing through the one-dimensional channel. 
Confirming this rough expectation, the superfluid weight $\ds$ as a function of disorder strength $W$
shown in Fig.\ref{fig:effectofhartree}a is strongly reduced when Hartree potentials are included.
However, an important caveat is that comparing the values with and without the Hartree term at the same value of bare disorder $W$ is not a fair comparison,
as the Hartree term effectively enhances the potential profile induced by the disorder.
Consequently, more insight into the true effect of the Hartree term can be obtained if we define 
an effective value of disorder using the full potential:
\begin{equation}
\label{eq:defrenormdisorder}
\tilde W^2 =\frac{1}{\nuc N_f} \sum_{\mathbf r,\slind}\disav{ (\epsilon^{Hartree}_{\mathbf r,\slind}+\epsilon_{\mathbf r,\slind})^2 },
\end{equation}
where $\epsilon_{\mathbf r,\slind}$ is the value of the disorder potential on site ${\mathbf r,\slind}$, and
$\epsilon^{Hartree}_{\mathbf r,\slind}$ is the corresponding value of the Hartree potential.
Defined in this way, the value of $\tilde W$ reflects the variance of the full potential given
as the sum of the bare disorder (of strength $W$) and the Hartree potential.
Figure~\ref{fig:effectofhartree}b shows the disordered superfluid weight as a function of $\tilde W$, showing it to be notably more robust to the addition of the Hartree term as a function of this effective disorder strength. 
Thus, a large part of the effect shown in Fig.\ref{fig:effectofhartree}a is due to the use of the bare disorder strength $W$.
The dependence of $\ds/\overline \Delta$ on $\tilde W$, shown in 
Fig.~\ref{fig:effectofhartree}c, brings out that this rescaled value is in fact robust
up to $\tilde W \approx 0.5 \overline \Delta \approx 0.13$, implying that our analytical theory (derived
neglecting Hartree effects) remains
valid for small disorder strengths $\tilde W$. 
In this sense, the effect of the Hartree term can be understood as increasing $W$ from the bare value,
and our analytical theory is valid so long as the effective potential is in the perturbative regime.

The reason why the superfluid weight in the Creutz ladder starts to decrease beyond the validity regime of our analytical results (above $\tilde W \approx 0.5 \overline \Delta$, e.g. for $\tilde W=0.13$) becomes apparent when considering the local filling at each site, $n_x$ (summed over spin and sublattice), shown
in Fig.~\ref{fig:effectofhartree}d.
This plot obtained for a single disorder realization shows that with the Hartree term, local disorder maxima produce regions that are almost depleted of electrons.
Consequently, the superfluid weight is greatly reduced, as for 1D superconductors connected in series,
it is the inverse of the superfluid weights that is additive.
Thus, a small region of near zero superfluid weight in the center (a weak link) can stop the supercurrent from flowing.

The effect of the Hartree term is markedly different in two dimensions: a single zero-density region would act merely as a local obstacle, around which supercurrent can flow. Indeed, a moving object that locally obstructs the superflow a typical probe of superfluidity. In two dimensions, superfluidity would thus be destroyed only when there are so many zero-density areas that the system breaks apart into isolated superconducting puddles. We show in Fig.~\ref{fig:effectofhartreelieb} that indeed for the two-dimensional Lieb lattice the effect of the Hartree term on the superfluid weight is much weaker than for the Creutz ladder, and the superfluid weight rescaled by the order parameter (Fig.~\ref{fig:effectofhartreelieb}c) is nearly unchanged by disorder.

Taken together, these numerical results show that our analytical results apply even when Hartree potentials are included, in particular for two-dimensional systems.
}

%% file: main.bbl
\begin{thebibliography}{45}%
\makeatletter
\providecommand \@ifxundefined [1]{%
 \@ifx{#1\undefined}
}%
\providecommand \@ifnum [1]{%
 \ifnum #1\expandafter \@firstoftwo
 \else \expandafter \@secondoftwo
 \fi
}%
\providecommand \@ifx [1]{%
 \ifx #1\expandafter \@firstoftwo
 \else \expandafter \@secondoftwo
 \fi
}%
\providecommand \natexlab [1]{#1}%
\providecommand \enquote  [1]{``#1''}%
\providecommand \bibnamefont  [1]{#1}%
\providecommand \bibfnamefont [1]{#1}%
\providecommand \citenamefont [1]{#1}%
\providecommand \href@noop [0]{\@secondoftwo}%
\providecommand \href [0]{\begingroup \@sanitize@url \@href}%
\providecommand \@href[1]{\@@startlink{#1}\@@href}%
\providecommand \@@href[1]{\endgroup#1\@@endlink}%
\providecommand \@sanitize@url [0]{\catcode `\\12\catcode `\$12\catcode `\&12\catcode `\#12\catcode `\^12\catcode `\_12\catcode `\%12\relax}%
\providecommand \@@startlink[1]{}%
\providecommand \@@endlink[0]{}%
\providecommand \url  [0]{\begingroup\@sanitize@url \@url }%
\providecommand \@url [1]{\endgroup\@href {#1}{\urlprefix }}%
\providecommand \urlprefix  [0]{URL }%
\providecommand \Eprint [0]{\href }%
\providecommand \doibase [0]{https://doi.org/}%
\providecommand \selectlanguage [0]{\@gobble}%
\providecommand \bibinfo  [0]{\@secondoftwo}%
\providecommand \bibfield  [0]{\@secondoftwo}%
\providecommand \translation [1]{[#1]}%
\providecommand \BibitemOpen [0]{}%
\providecommand \bibitemStop [0]{}%
\providecommand \bibitemNoStop [0]{.\EOS\space}%
\providecommand \EOS [0]{\spacefactor3000\relax}%
\providecommand \BibitemShut  [1]{\csname bibitem#1\endcsname}%
\let\auto@bib@innerbib\@empty
\bibitem [{\citenamefont {Cao}\ \emph {et~al.}(2018{\natexlab{a}})\citenamefont {Cao}, \citenamefont {Fatemi}, \citenamefont {Fang}, \citenamefont {Watanabe}, \citenamefont {Taniguchi}, \citenamefont {Kaxiras},\ and\ \citenamefont {Jarillo-Herrero}}]{jarillo-herreroCaoUnconventionalSuperconductivityMagicangle2018}%
  \BibitemOpen
  \bibfield  {author} {\bibinfo {author} {\bibfnamefont {Y.}~\bibnamefont {Cao}}, \bibinfo {author} {\bibfnamefont {V.}~\bibnamefont {Fatemi}}, \bibinfo {author} {\bibfnamefont {S.}~\bibnamefont {Fang}}, \bibinfo {author} {\bibfnamefont {K.}~\bibnamefont {Watanabe}}, \bibinfo {author} {\bibfnamefont {T.}~\bibnamefont {Taniguchi}}, \bibinfo {author} {\bibfnamefont {E.}~\bibnamefont {Kaxiras}},\ and\ \bibinfo {author} {\bibfnamefont {P.}~\bibnamefont {Jarillo-Herrero}},\ }\bibfield  {title} {\bibinfo {title} {Unconventional superconductivity in magic-angle graphene superlattices},\ }\href {https://doi.org/10.1038/nature26160} {\bibfield  {journal} {\bibinfo  {journal} {Nature}\ }\textbf {\bibinfo {volume} {556}},\ \bibinfo {pages} {43} (\bibinfo {year} {2018}{\natexlab{a}})}\BibitemShut {NoStop}%
\bibitem [{\citenamefont {Cao}\ \emph {et~al.}(2018{\natexlab{b}})\citenamefont {Cao}, \citenamefont {Fatemi}, \citenamefont {Demir}, \citenamefont {Fang}, \citenamefont {Tomarken}, \citenamefont {Luo}, \citenamefont {Sanchez-Yamagishi}, \citenamefont {Watanabe}, \citenamefont {Taniguchi}, \citenamefont {Kaxiras}, \citenamefont {Ashoori},\ and\ \citenamefont {Jarillo-Herrero}}]{jarillo-herreroCaoCorrelatedInsulatorBehaviour2018}%
  \BibitemOpen
  \bibfield  {author} {\bibinfo {author} {\bibfnamefont {Y.}~\bibnamefont {Cao}}, \bibinfo {author} {\bibfnamefont {V.}~\bibnamefont {Fatemi}}, \bibinfo {author} {\bibfnamefont {A.}~\bibnamefont {Demir}}, \bibinfo {author} {\bibfnamefont {S.}~\bibnamefont {Fang}}, \bibinfo {author} {\bibfnamefont {S.~L.}\ \bibnamefont {Tomarken}}, \bibinfo {author} {\bibfnamefont {J.~Y.}\ \bibnamefont {Luo}}, \bibinfo {author} {\bibfnamefont {J.~D.}\ \bibnamefont {Sanchez-Yamagishi}}, \bibinfo {author} {\bibfnamefont {K.}~\bibnamefont {Watanabe}}, \bibinfo {author} {\bibfnamefont {T.}~\bibnamefont {Taniguchi}}, \bibinfo {author} {\bibfnamefont {E.}~\bibnamefont {Kaxiras}}, \bibinfo {author} {\bibfnamefont {R.~C.}\ \bibnamefont {Ashoori}},\ and\ \bibinfo {author} {\bibfnamefont {P.}~\bibnamefont {Jarillo-Herrero}},\ }\bibfield  {title} {\bibinfo {title} {Correlated insulator behaviour at half-filling in magic-angle graphene superlattices},\ }\href {https://doi.org/10.1038/nature26154} {\bibfield  {journal} {\bibinfo  {journal} {Nature}\ }\textbf {\bibinfo {volume} {556}},\ \bibinfo {pages} {80} (\bibinfo {year} {2018}{\natexlab{b}})}\BibitemShut {NoStop}%
\bibitem [{\citenamefont {Balents}\ \emph {et~al.}(2020)\citenamefont {Balents}, \citenamefont {Dean}, \citenamefont {Efetov},\ and\ \citenamefont {Young}}]{youngBalentsSuperconductivityStrongCorrelations2020}%
  \BibitemOpen
  \bibfield  {author} {\bibinfo {author} {\bibfnamefont {L.}~\bibnamefont {Balents}}, \bibinfo {author} {\bibfnamefont {C.~R.}\ \bibnamefont {Dean}}, \bibinfo {author} {\bibfnamefont {D.~K.}\ \bibnamefont {Efetov}},\ and\ \bibinfo {author} {\bibfnamefont {A.~F.}\ \bibnamefont {Young}},\ }\bibfield  {title} {\bibinfo {title} {Superconductivity and strong correlations in moir{\'e} flat bands},\ }\href {https://doi.org/10.1038/s41567-020-0906-9} {\bibfield  {journal} {\bibinfo  {journal} {Nature Physics}\ }\textbf {\bibinfo {volume} {16}},\ \bibinfo {pages} {725} (\bibinfo {year} {2020})}\BibitemShut {NoStop}%
\bibitem [{\citenamefont {Kennes}\ \emph {et~al.}(2021)\citenamefont {Kennes}, \citenamefont {Claassen}, \citenamefont {Xian}, \citenamefont {Georges}, \citenamefont {Millis}, \citenamefont {Hone}, \citenamefont {Dean}, \citenamefont {Basov}, \citenamefont {Pasupathy},\ and\ \citenamefont {Rubio}}]{rubioKennesMoireHeterostructuresCondensedmatter2021}%
  \BibitemOpen
  \bibfield  {author} {\bibinfo {author} {\bibfnamefont {D.~M.}\ \bibnamefont {Kennes}}, \bibinfo {author} {\bibfnamefont {M.}~\bibnamefont {Claassen}}, \bibinfo {author} {\bibfnamefont {L.}~\bibnamefont {Xian}}, \bibinfo {author} {\bibfnamefont {A.}~\bibnamefont {Georges}}, \bibinfo {author} {\bibfnamefont {A.~J.}\ \bibnamefont {Millis}}, \bibinfo {author} {\bibfnamefont {J.}~\bibnamefont {Hone}}, \bibinfo {author} {\bibfnamefont {C.~R.}\ \bibnamefont {Dean}}, \bibinfo {author} {\bibfnamefont {D.~N.}\ \bibnamefont {Basov}}, \bibinfo {author} {\bibfnamefont {A.~N.}\ \bibnamefont {Pasupathy}},\ and\ \bibinfo {author} {\bibfnamefont {A.}~\bibnamefont {Rubio}},\ }\bibfield  {title} {\bibinfo {title} {Moiré heterostructures as a condensed-matter quantum simulator},\ }\href {https://doi.org/10.1038/s41567-020-01154-3} {\bibfield  {journal} {\bibinfo  {journal} {Nature Physics}\ }\textbf {\bibinfo {volume} {17}},\ \bibinfo {pages} {155} (\bibinfo {year} {2021})}\BibitemShut {NoStop}%
\bibitem [{\citenamefont {Andrei}\ \emph {et~al.}(2021)\citenamefont {Andrei}, \citenamefont {Efetov}, \citenamefont {Jarillo-Herrero}, \citenamefont {MacDonald}, \citenamefont {Mak}, \citenamefont {Senthil}, \citenamefont {Tutuc}, \citenamefont {Yazdani},\ and\ \citenamefont {Young}}]{youngAndreiMarvelsMoireMaterials2021}%
  \BibitemOpen
  \bibfield  {author} {\bibinfo {author} {\bibfnamefont {E.~Y.}\ \bibnamefont {Andrei}}, \bibinfo {author} {\bibfnamefont {D.~K.}\ \bibnamefont {Efetov}}, \bibinfo {author} {\bibfnamefont {P.}~\bibnamefont {Jarillo-Herrero}}, \bibinfo {author} {\bibfnamefont {A.~H.}\ \bibnamefont {MacDonald}}, \bibinfo {author} {\bibfnamefont {K.~F.}\ \bibnamefont {Mak}}, \bibinfo {author} {\bibfnamefont {T.}~\bibnamefont {Senthil}}, \bibinfo {author} {\bibfnamefont {E.}~\bibnamefont {Tutuc}}, \bibinfo {author} {\bibfnamefont {A.}~\bibnamefont {Yazdani}},\ and\ \bibinfo {author} {\bibfnamefont {A.~F.}\ \bibnamefont {Young}},\ }\bibfield  {title} {\bibinfo {title} {The marvels of moiré materials},\ }\href {https://doi.org/10.1038/s41578-021-00284-1} {\bibfield  {journal} {\bibinfo  {journal} {Nature Reviews Materials}\ }\textbf {\bibinfo {volume} {6}},\ \bibinfo {pages} {201} (\bibinfo {year} {2021})}\BibitemShut {NoStop}%
\bibitem [{\citenamefont {Törmä}\ \emph {et~al.}(2022)\citenamefont {Törmä}, \citenamefont {Peotta},\ and\ \citenamefont {Bernevig}}]{bernevigTormaSuperconductivitySuperfluidityQuantum2022}%
  \BibitemOpen
  \bibfield  {author} {\bibinfo {author} {\bibfnamefont {P.}~\bibnamefont {Törmä}}, \bibinfo {author} {\bibfnamefont {S.}~\bibnamefont {Peotta}},\ and\ \bibinfo {author} {\bibfnamefont {B.~A.}\ \bibnamefont {Bernevig}},\ }\bibfield  {title} {\bibinfo {title} {Superconductivity, superfluidity and quantum geometry in twisted multilayer systems},\ }\href {https://doi.org/10.1038/s42254-022-00466-y} {\bibfield  {journal} {\bibinfo  {journal} {Nature Reviews Physics}\ }\textbf {\bibinfo {volume} {4}},\ \bibinfo {pages} {528} (\bibinfo {year} {2022})}\BibitemShut {NoStop}%
\bibitem [{\citenamefont {Tanaka}\ \emph {et~al.}(2025)\citenamefont {Tanaka}, \citenamefont {Wang}, \citenamefont {Dinh}, \citenamefont {{Rodan-Legrain}}, \citenamefont {Zaman}, \citenamefont {Hays}, \citenamefont {Almanakly}, \citenamefont {Kannan}, \citenamefont {Kim}, \citenamefont {Niedzielski}, \citenamefont {Serniak}, \citenamefont {Schwartz}, \citenamefont {Watanabe}, \citenamefont {Taniguchi}, \citenamefont {Orlando}, \citenamefont {Gustavsson}, \citenamefont {Grover}, \citenamefont {{Jarillo-Herrero}},\ and\ \citenamefont {Oliver}}]{oliverTanakaSuperfluidStiffnessMagicangle2025}%
  \BibitemOpen
  \bibfield  {author} {\bibinfo {author} {\bibfnamefont {M.}~\bibnamefont {Tanaka}}, \bibinfo {author} {\bibfnamefont {J.~{\^I}.-j.}\ \bibnamefont {Wang}}, \bibinfo {author} {\bibfnamefont {T.~H.}\ \bibnamefont {Dinh}}, \bibinfo {author} {\bibfnamefont {D.}~\bibnamefont {{Rodan-Legrain}}}, \bibinfo {author} {\bibfnamefont {S.}~\bibnamefont {Zaman}}, \bibinfo {author} {\bibfnamefont {M.}~\bibnamefont {Hays}}, \bibinfo {author} {\bibfnamefont {A.}~\bibnamefont {Almanakly}}, \bibinfo {author} {\bibfnamefont {B.}~\bibnamefont {Kannan}}, \bibinfo {author} {\bibfnamefont {D.~K.}\ \bibnamefont {Kim}}, \bibinfo {author} {\bibfnamefont {B.~M.}\ \bibnamefont {Niedzielski}}, \bibinfo {author} {\bibfnamefont {K.}~\bibnamefont {Serniak}}, \bibinfo {author} {\bibfnamefont {M.~E.}\ \bibnamefont {Schwartz}}, \bibinfo {author} {\bibfnamefont {K.}~\bibnamefont {Watanabe}}, \bibinfo {author} {\bibfnamefont {T.}~\bibnamefont {Taniguchi}}, \bibinfo {author} {\bibfnamefont {T.~P.}\ \bibnamefont {Orlando}}, \bibinfo {author} {\bibfnamefont {S.}~\bibnamefont {Gustavsson}}, \bibinfo {author} {\bibfnamefont {J.~A.}\ \bibnamefont {Grover}}, \bibinfo {author} {\bibfnamefont {P.}~\bibnamefont {{Jarillo-Herrero}}},\ and\ \bibinfo {author} {\bibfnamefont {W.~D.}\ \bibnamefont {Oliver}},\ }\bibfield  {title} {\bibinfo {title} {Superfluid stiffness of magic-angle twisted bilayer graphene},\ }\href {https://doi.org/10.1038/s41586-024-08494-7} {\bibfield  {journal} {\bibinfo  {journal} {Nature}\ }\textbf {\bibinfo {volume} {638}},\ \bibinfo {pages} {99} (\bibinfo {year} {2025})}\BibitemShut {NoStop}%
\bibitem [{\citenamefont {Tian}\ \emph {et~al.}(2023)\citenamefont {Tian}, \citenamefont {Gao}, \citenamefont {Zhang}, \citenamefont {Che}, \citenamefont {Xu}, \citenamefont {Cheung}, \citenamefont {Watanabe}, \citenamefont {Taniguchi}, \citenamefont {Randeria}, \citenamefont {Zhang}, \citenamefont {Lau},\ and\ \citenamefont {Bockrath}}]{bockrathTianEvidenceDiracFlat2023}%
  \BibitemOpen
  \bibfield  {author} {\bibinfo {author} {\bibfnamefont {H.}~\bibnamefont {Tian}}, \bibinfo {author} {\bibfnamefont {X.}~\bibnamefont {Gao}}, \bibinfo {author} {\bibfnamefont {Y.}~\bibnamefont {Zhang}}, \bibinfo {author} {\bibfnamefont {S.}~\bibnamefont {Che}}, \bibinfo {author} {\bibfnamefont {T.}~\bibnamefont {Xu}}, \bibinfo {author} {\bibfnamefont {P.}~\bibnamefont {Cheung}}, \bibinfo {author} {\bibfnamefont {K.}~\bibnamefont {Watanabe}}, \bibinfo {author} {\bibfnamefont {T.}~\bibnamefont {Taniguchi}}, \bibinfo {author} {\bibfnamefont {M.}~\bibnamefont {Randeria}}, \bibinfo {author} {\bibfnamefont {F.}~\bibnamefont {Zhang}}, \bibinfo {author} {\bibfnamefont {C.~N.}\ \bibnamefont {Lau}},\ and\ \bibinfo {author} {\bibfnamefont {M.~W.}\ \bibnamefont {Bockrath}},\ }\bibfield  {title} {\bibinfo {title} {Evidence for dirac flat band superconductivity enabled by quantum geometry},\ }\href {https://doi.org/10.1038/s41586-022-05576-2} {\bibfield  {journal} {\bibinfo  {journal} {Nature}\ }\textbf {\bibinfo {volume} {614}},\ \bibinfo {pages} {440} (\bibinfo {year} {2023})}\BibitemShut {NoStop}%
\bibitem [{\citenamefont {Zhou}\ \emph {et~al.}(2021{\natexlab{a}})\citenamefont {Zhou}, \citenamefont {Xie}, \citenamefont {Taniguchi}, \citenamefont {Watanabe},\ and\ \citenamefont {Young}}]{youngZhouSuperconductivityRhombohedralTrilayer2021}%
  \BibitemOpen
  \bibfield  {author} {\bibinfo {author} {\bibfnamefont {H.}~\bibnamefont {Zhou}}, \bibinfo {author} {\bibfnamefont {T.}~\bibnamefont {Xie}}, \bibinfo {author} {\bibfnamefont {T.}~\bibnamefont {Taniguchi}}, \bibinfo {author} {\bibfnamefont {K.}~\bibnamefont {Watanabe}},\ and\ \bibinfo {author} {\bibfnamefont {A.~F.}\ \bibnamefont {Young}},\ }\bibfield  {title} {\bibinfo {title} {Superconductivity in rhombohedral trilayer graphene},\ }\href {https://doi.org/10.1038/s41586-021-03926-0} {\bibfield  {journal} {\bibinfo  {journal} {Nature}\ }\textbf {\bibinfo {volume} {598}},\ \bibinfo {pages} {434} (\bibinfo {year} {2021}{\natexlab{a}})}\BibitemShut {NoStop}%
\bibitem [{\citenamefont {Zhou}\ \emph {et~al.}(2021{\natexlab{b}})\citenamefont {Zhou}, \citenamefont {Xie}, \citenamefont {Ghazaryan}, \citenamefont {Holder}, \citenamefont {Ehrets}, \citenamefont {Spanton}, \citenamefont {Taniguchi}, \citenamefont {Watanabe}, \citenamefont {Berg}, \citenamefont {Serbyn},\ and\ \citenamefont {Young}}]{youngZhouHalfQuartermetalsRhombohedral2021}%
  \BibitemOpen
  \bibfield  {author} {\bibinfo {author} {\bibfnamefont {H.}~\bibnamefont {Zhou}}, \bibinfo {author} {\bibfnamefont {T.}~\bibnamefont {Xie}}, \bibinfo {author} {\bibfnamefont {A.}~\bibnamefont {Ghazaryan}}, \bibinfo {author} {\bibfnamefont {T.}~\bibnamefont {Holder}}, \bibinfo {author} {\bibfnamefont {J.~R.}\ \bibnamefont {Ehrets}}, \bibinfo {author} {\bibfnamefont {E.~M.}\ \bibnamefont {Spanton}}, \bibinfo {author} {\bibfnamefont {T.}~\bibnamefont {Taniguchi}}, \bibinfo {author} {\bibfnamefont {K.}~\bibnamefont {Watanabe}}, \bibinfo {author} {\bibfnamefont {E.}~\bibnamefont {Berg}}, \bibinfo {author} {\bibfnamefont {M.}~\bibnamefont {Serbyn}},\ and\ \bibinfo {author} {\bibfnamefont {A.~F.}\ \bibnamefont {Young}},\ }\bibfield  {title} {\bibinfo {title} {Half- and quarter-metals in rhombohedral trilayer graphene},\ }\href {https://doi.org/10.1038/s41586-021-03938-w} {\bibfield  {journal} {\bibinfo  {journal} {Nature}\ }\textbf {\bibinfo {volume} {598}},\ \bibinfo {pages} {429} (\bibinfo {year} {2021}{\natexlab{b}})}\BibitemShut {NoStop}%
\bibitem [{\citenamefont {Lu}\ \emph {et~al.}(2024)\citenamefont {Lu}, \citenamefont {Han}, \citenamefont {Yao}, \citenamefont {Reddy}, \citenamefont {Yang}, \citenamefont {Seo}, \citenamefont {Watanabe}, \citenamefont {Taniguchi}, \citenamefont {Fu},\ and\ \citenamefont {Ju}}]{juLuFractionalQuantumAnomalous2024}%
  \BibitemOpen
  \bibfield  {author} {\bibinfo {author} {\bibfnamefont {Z.}~\bibnamefont {Lu}}, \bibinfo {author} {\bibfnamefont {T.}~\bibnamefont {Han}}, \bibinfo {author} {\bibfnamefont {Y.}~\bibnamefont {Yao}}, \bibinfo {author} {\bibfnamefont {A.~P.}\ \bibnamefont {Reddy}}, \bibinfo {author} {\bibfnamefont {J.}~\bibnamefont {Yang}}, \bibinfo {author} {\bibfnamefont {J.}~\bibnamefont {Seo}}, \bibinfo {author} {\bibfnamefont {K.}~\bibnamefont {Watanabe}}, \bibinfo {author} {\bibfnamefont {T.}~\bibnamefont {Taniguchi}}, \bibinfo {author} {\bibfnamefont {L.}~\bibnamefont {Fu}},\ and\ \bibinfo {author} {\bibfnamefont {L.}~\bibnamefont {Ju}},\ }\bibfield  {title} {\bibinfo {title} {Fractional quantum anomalous hall effect in multilayer graphene},\ }\href {https://doi.org/10.1038/s41586-023-07010-7} {\bibfield  {journal} {\bibinfo  {journal} {Nature}\ }\textbf {\bibinfo {volume} {626}},\ \bibinfo {pages} {759} (\bibinfo {year} {2024})}\BibitemShut {NoStop}%
\bibitem [{\citenamefont {Leykam}\ \emph {et~al.}(2018)\citenamefont {Leykam}, \citenamefont {Andreanov},\ and\ \citenamefont {Flach}}]{flachLeykamArtificialFlatBand2018}%
  \BibitemOpen
  \bibfield  {author} {\bibinfo {author} {\bibfnamefont {D.}~\bibnamefont {Leykam}}, \bibinfo {author} {\bibfnamefont {A.}~\bibnamefont {Andreanov}},\ and\ \bibinfo {author} {\bibfnamefont {S.}~\bibnamefont {Flach}},\ }\bibfield  {title} {\bibinfo {title} {Artificial flat band systems: from lattice models to experiments},\ }\href {https://doi.org/10.1080/23746149.2018.1473052} {\bibfield  {journal} {\bibinfo  {journal} {Advances in Physics: X}\ }\textbf {\bibinfo {volume} {3}},\ \bibinfo {pages} {1473052} (\bibinfo {year} {2018})}\BibitemShut {NoStop}%
\bibitem [{\citenamefont {Peotta}\ and\ \citenamefont {T{\"o}rm{\"a}}(2015)}]{tormaPeottaSuperfluidityTopologicallyNontrivial2015}%
  \BibitemOpen
  \bibfield  {author} {\bibinfo {author} {\bibfnamefont {S.}~\bibnamefont {Peotta}}\ and\ \bibinfo {author} {\bibfnamefont {P.}~\bibnamefont {T{\"o}rm{\"a}}},\ }\bibfield  {title} {\bibinfo {title} {Superfluidity in topologically nontrivial flat bands},\ }\href {https://doi.org/10.1038/ncomms9944} {\bibfield  {journal} {\bibinfo  {journal} {Nature Communications}\ }\textbf {\bibinfo {volume} {6}},\ \bibinfo {pages} {8944} (\bibinfo {year} {2015})}\BibitemShut {NoStop}%
\bibitem [{\citenamefont {Tovmasyan}\ \emph {et~al.}(2018)\citenamefont {Tovmasyan}, \citenamefont {Peotta}, \citenamefont {Liang}, \citenamefont {T{\"o}rm{\"a}},\ and\ \citenamefont {Huber}}]{huberTovmasyanPreformedPairsFlat2018}%
  \BibitemOpen
  \bibfield  {author} {\bibinfo {author} {\bibfnamefont {M.}~\bibnamefont {Tovmasyan}}, \bibinfo {author} {\bibfnamefont {S.}~\bibnamefont {Peotta}}, \bibinfo {author} {\bibfnamefont {L.}~\bibnamefont {Liang}}, \bibinfo {author} {\bibfnamefont {P.}~\bibnamefont {T{\"o}rm{\"a}}},\ and\ \bibinfo {author} {\bibfnamefont {S.~D.}\ \bibnamefont {Huber}},\ }\bibfield  {title} {\bibinfo {title} {Preformed pairs in flat {{Bloch}} bands},\ }\href {https://doi.org/10.1103/PhysRevB.98.134513} {\bibfield  {journal} {\bibinfo  {journal} {Physical Review B}\ }\textbf {\bibinfo {volume} {98}},\ \bibinfo {pages} {134513} (\bibinfo {year} {2018})}\BibitemShut {NoStop}%
\bibitem [{\citenamefont {T{\"o}rm{\"a}}\ \emph {et~al.}(2018)\citenamefont {T{\"o}rm{\"a}}, \citenamefont {Liang},\ and\ \citenamefont {Peotta}}]{peottaTormaQuantumMetricEffective2018}%
  \BibitemOpen
  \bibfield  {author} {\bibinfo {author} {\bibfnamefont {P.}~\bibnamefont {T{\"o}rm{\"a}}}, \bibinfo {author} {\bibfnamefont {L.}~\bibnamefont {Liang}},\ and\ \bibinfo {author} {\bibfnamefont {S.}~\bibnamefont {Peotta}},\ }\bibfield  {title} {\bibinfo {title} {Quantum metric and effective mass of a two-body bound state in a flat band},\ }\href {https://doi.org/10.1103/PhysRevB.98.220511} {\bibfield  {journal} {\bibinfo  {journal} {Physical Review B}\ }\textbf {\bibinfo {volume} {98}},\ \bibinfo {pages} {220511} (\bibinfo {year} {2018})}\BibitemShut {NoStop}%
\bibitem [{\citenamefont {Huhtinen}\ \emph {et~al.}(2022)\citenamefont {Huhtinen}, \citenamefont {Herzog-Arbeitman}, \citenamefont {Chew}, \citenamefont {Bernevig},\ and\ \citenamefont {Törmä}}]{tormaHuhtinenRevisitingFlatBand2022}%
  \BibitemOpen
  \bibfield  {author} {\bibinfo {author} {\bibfnamefont {K.-E.}\ \bibnamefont {Huhtinen}}, \bibinfo {author} {\bibfnamefont {J.}~\bibnamefont {Herzog-Arbeitman}}, \bibinfo {author} {\bibfnamefont {A.}~\bibnamefont {Chew}}, \bibinfo {author} {\bibfnamefont {B.~A.}\ \bibnamefont {Bernevig}},\ and\ \bibinfo {author} {\bibfnamefont {P.}~\bibnamefont {Törmä}},\ }\bibfield  {title} {\bibinfo {title} {Revisiting flat band superconductivity: Dependence on minimal quantum metric and band touchings},\ }\href {https://doi.org/10.1103/PhysRevB.106.014518} {\bibfield  {journal} {\bibinfo  {journal} {Physical Review B}\ }\textbf {\bibinfo {volume} {106}},\ \bibinfo {pages} {014518} (\bibinfo {year} {2022})}\BibitemShut {NoStop}%
\bibitem [{\citenamefont {Lau}\ \emph {et~al.}(2022)\citenamefont {Lau}, \citenamefont {Peotta}, \citenamefont {Pikulin}, \citenamefont {Rossi},\ and\ \citenamefont {Hyart}}]{hyartLauUniversalSuppressionSuperfluid2022}%
  \BibitemOpen
  \bibfield  {author} {\bibinfo {author} {\bibfnamefont {A.}~\bibnamefont {Lau}}, \bibinfo {author} {\bibfnamefont {S.}~\bibnamefont {Peotta}}, \bibinfo {author} {\bibfnamefont {D.}~\bibnamefont {Pikulin}}, \bibinfo {author} {\bibfnamefont {E.}~\bibnamefont {Rossi}},\ and\ \bibinfo {author} {\bibfnamefont {T.}~\bibnamefont {Hyart}},\ }\bibfield  {title} {\bibinfo {title} {Universal suppression of superfluid weight by non-magnetic disorder in $s$-wave superconductors independent of quantum geometry and band dispersion},\ }\href {https://doi.org/10.21468/SciPostPhys.13.4.086} {\bibfield  {journal} {\bibinfo  {journal} {SciPost Physics}\ }\textbf {\bibinfo {volume} {13}},\ \bibinfo {pages} {086} (\bibinfo {year} {2022})}\BibitemShut {NoStop}%
\bibitem [{\citenamefont {Liang}\ \emph {et~al.}(2023)\citenamefont {Liang}, \citenamefont {Yang}, \citenamefont {Cheng},\ and\ \citenamefont {Mondaini}}]{mondainiLiangDisorderInteractingQuasionedimensional2023}%
  \BibitemOpen
  \bibfield  {author} {\bibinfo {author} {\bibfnamefont {M.-J.}\ \bibnamefont {Liang}}, \bibinfo {author} {\bibfnamefont {Y.-F.}\ \bibnamefont {Yang}}, \bibinfo {author} {\bibfnamefont {C.}~\bibnamefont {Cheng}},\ and\ \bibinfo {author} {\bibfnamefont {R.}~\bibnamefont {Mondaini}},\ }\bibfield  {title} {\bibinfo {title} {Disorder in interacting quasi-one-dimensional systems: Flat and dispersive bands},\ }\href {https://doi.org/10.1103/PhysRevB.108.035131} {\bibfield  {journal} {\bibinfo  {journal} {Physical Review B}\ }\textbf {\bibinfo {volume} {108}},\ \bibinfo {pages} {035131} (\bibinfo {year} {2023})}\BibitemShut {NoStop}%
\bibitem [{\citenamefont {Chan}\ \emph {et~al.}(2025)\citenamefont {Chan}, \citenamefont {Grémaud},\ and\ \citenamefont {Batrouni}}]{batrouniChanDisorderRobustnessSuperconductivity2025}%
  \BibitemOpen
  \bibfield  {author} {\bibinfo {author} {\bibfnamefont {S.~M.}\ \bibnamefont {Chan}}, \bibinfo {author} {\bibfnamefont {B.}~\bibnamefont {Grémaud}},\ and\ \bibinfo {author} {\bibfnamefont {G.~G.}\ \bibnamefont {Batrouni}},\ }\href {https://doi.org/10.48550/arXiv.2506.07095} {\bibinfo {title} {Disorder and the robustness of superconductivity on the flat band}} (\bibinfo {year} {2025}),\ \Eprint {https://arxiv.org/abs/2506.07095} {arXiv:2506.07095 [cond-mat]} \BibitemShut {NoStop}%
\bibitem [{\citenamefont {Bouzerar}\ and\ \citenamefont {Thumin}(2025)}]{thuminBouzerarRobustnessFlatBand2025}%
  \BibitemOpen
  \bibfield  {author} {\bibinfo {author} {\bibfnamefont {G.}~\bibnamefont {Bouzerar}}\ and\ \bibinfo {author} {\bibfnamefont {M.}~\bibnamefont {Thumin}},\ }\bibfield  {title} {\bibinfo {title} {Robustness of flat band superconductivity against disorder in a two-dimensional {{Lieb}} lattice model},\ }\href {https://doi.org/10.1103/PhysRevB.111.L020506} {\bibfield  {journal} {\bibinfo  {journal} {Physical Review B}\ }\textbf {\bibinfo {volume} {111}},\ \bibinfo {pages} {L020506} (\bibinfo {year} {2025})}\BibitemShut {NoStop}%
\bibitem [{\citenamefont {Marzari}\ and\ \citenamefont {Vanderbilt}(1997)}]{vanderbiltMarzariMaximallyLocalizedGeneralized1997}%
  \BibitemOpen
  \bibfield  {author} {\bibinfo {author} {\bibfnamefont {N.}~\bibnamefont {Marzari}}\ and\ \bibinfo {author} {\bibfnamefont {D.}~\bibnamefont {Vanderbilt}},\ }\bibfield  {title} {\bibinfo {title} {Maximally localized generalized wannier functions for composite energy bands},\ }\href {https://doi.org/10.1103/PhysRevB.56.12847} {\bibfield  {journal} {\bibinfo  {journal} {Physical Review B}\ }\textbf {\bibinfo {volume} {56}},\ \bibinfo {pages} {12847} (\bibinfo {year} {1997})}\BibitemShut {NoStop}%
\bibitem [{\citenamefont {Yu}\ \emph {et~al.}(2024)\citenamefont {Yu}, \citenamefont {Bernevig}, \citenamefont {Queiroz}, \citenamefont {Rossi}, \citenamefont {T{\"o}rm{\"a}},\ and\ \citenamefont {Yang}}]{yangYuQuantumGeometryQuantum2024}%
  \BibitemOpen
  \bibfield  {author} {\bibinfo {author} {\bibfnamefont {J.}~\bibnamefont {Yu}}, \bibinfo {author} {\bibfnamefont {B.~A.}\ \bibnamefont {Bernevig}}, \bibinfo {author} {\bibfnamefont {R.}~\bibnamefont {Queiroz}}, \bibinfo {author} {\bibfnamefont {E.}~\bibnamefont {Rossi}}, \bibinfo {author} {\bibfnamefont {P.}~\bibnamefont {T{\"o}rm{\"a}}},\ and\ \bibinfo {author} {\bibfnamefont {B.-J.}\ \bibnamefont {Yang}},\ }\href {https://doi.org/10.48550/arXiv.2501.00098} {\bibinfo {title} {Quantum {{Geometry}} in {{Quantum Materials}}}} (\bibinfo {year} {2024}),\ \Eprint {https://arxiv.org/abs/2501.00098} {arXiv:2501.00098 [cond-mat]} \BibitemShut {NoStop}%
\bibitem [{sup()}]{supplement}%
  \BibitemOpen
  \href@noop {} {}\bibinfo {note} {See Supplementary Material for details on the models used, numerical methods, and a derivation of the equivalence of projected and unprojected disorder. It includes Refs.~\cite{vanderbiltMarzariMaximallyLocalizedWannier2012,tormaHuhtinenRevisitingFlatBand2022,lebrisCancesCanWeOutperform2000,tormaPeottaSuperfluidityTopologicallyNontrivial2015}}\BibitemShut {NoStop}%
\bibitem [{\citenamefont {Ghosal}\ \emph {et~al.}(1998)\citenamefont {Ghosal}, \citenamefont {Randeria},\ and\ \citenamefont {Trivedi}}]{trivediGhosalRoleSpatialAmplitude1998}%
  \BibitemOpen
  \bibfield  {author} {\bibinfo {author} {\bibfnamefont {A.}~\bibnamefont {Ghosal}}, \bibinfo {author} {\bibfnamefont {M.}~\bibnamefont {Randeria}},\ and\ \bibinfo {author} {\bibfnamefont {N.}~\bibnamefont {Trivedi}},\ }\bibfield  {title} {\bibinfo {title} {Role of spatial amplitude fluctuations in highly disordered $\mathit{s}$-wave superconductors},\ }\href {https://doi.org/10.1103/PhysRevLett.81.3940} {\bibfield  {journal} {\bibinfo  {journal} {Physical Review Letters}\ }\textbf {\bibinfo {volume} {81}},\ \bibinfo {pages} {3940} (\bibinfo {year} {1998})}\BibitemShut {NoStop}%
\bibitem [{\citenamefont {{\v C}ade{\v z}}\ \emph {et~al.}(2021)\citenamefont {{\v C}ade{\v z}}, \citenamefont {Kim}, \citenamefont {Andreanov},\ and\ \citenamefont {Flach}}]{flachCadezMetalinsulatorTransitionInfinitesimally2021}%
  \BibitemOpen
  \bibfield  {author} {\bibinfo {author} {\bibfnamefont {T.}~\bibnamefont {{\v C}ade{\v z}}}, \bibinfo {author} {\bibfnamefont {Y.}~\bibnamefont {Kim}}, \bibinfo {author} {\bibfnamefont {A.}~\bibnamefont {Andreanov}},\ and\ \bibinfo {author} {\bibfnamefont {S.}~\bibnamefont {Flach}},\ }\bibfield  {title} {\bibinfo {title} {Metal-insulator transition in infinitesimally weakly disordered flat bands},\ }\href {https://doi.org/10.1103/PhysRevB.104.L180201} {\bibfield  {journal} {\bibinfo  {journal} {Physical Review B}\ }\textbf {\bibinfo {volume} {104}},\ \bibinfo {pages} {L180201} (\bibinfo {year} {2021})}\BibitemShut {NoStop}%
\bibitem [{\citenamefont {Chau}\ \emph {et~al.}(2024)\citenamefont {Chau}, \citenamefont {Xiang}, \citenamefont {Chen},\ and\ \citenamefont {Law}}]{lawChauDisorderinducedDelocalizationFlatband2024}%
  \BibitemOpen
  \bibfield  {author} {\bibinfo {author} {\bibfnamefont {C.~W.}\ \bibnamefont {Chau}}, \bibinfo {author} {\bibfnamefont {T.}~\bibnamefont {Xiang}}, \bibinfo {author} {\bibfnamefont {S.~A.}\ \bibnamefont {Chen}},\ and\ \bibinfo {author} {\bibfnamefont {K.~T.}\ \bibnamefont {Law}},\ }\href {https://doi.org/10.48550/arXiv.2412.19056} {\bibinfo {title} {Disorder-induced delocalization in flat-band systems with quantum geometry}} (\bibinfo {year} {2024}),\ \Eprint {https://arxiv.org/abs/2412.19056} {arXiv:2412.19056 [cond-mat]} \BibitemShut {NoStop}%
\bibitem [{\citenamefont {Tovmasyan}\ \emph {et~al.}(2016)\citenamefont {Tovmasyan}, \citenamefont {Peotta}, \citenamefont {T{\"o}rm{\"a}},\ and\ \citenamefont {Huber}}]{huberTovmasyanEffectiveTheoryEmergent2016}%
  \BibitemOpen
  \bibfield  {author} {\bibinfo {author} {\bibfnamefont {M.}~\bibnamefont {Tovmasyan}}, \bibinfo {author} {\bibfnamefont {S.}~\bibnamefont {Peotta}}, \bibinfo {author} {\bibfnamefont {P.}~\bibnamefont {T{\"o}rm{\"a}}},\ and\ \bibinfo {author} {\bibfnamefont {S.~D.}\ \bibnamefont {Huber}},\ }\bibfield  {title} {\bibinfo {title} {Effective theory and emergent {{SU}} ( 2 ) symmetry in the flat bands of attractive {{Hubbard}} models},\ }\href {https://doi.org/10.1103/PhysRevB.94.245149} {\bibfield  {journal} {\bibinfo  {journal} {Physical Review B}\ }\textbf {\bibinfo {volume} {94}},\ \bibinfo {pages} {245149} (\bibinfo {year} {2016})}\BibitemShut {NoStop}%
\bibitem [{\citenamefont {Herzog-Arbeitman}\ \emph {et~al.}(2022)\citenamefont {Herzog-Arbeitman}, \citenamefont {Chew}, \citenamefont {Huhtinen}, \citenamefont {Törmä},\ and\ \citenamefont {Bernevig}}]{bernevigHerzog-ArbeitmanManyBodySuperconductivityTopological2022}%
  \BibitemOpen
  \bibfield  {author} {\bibinfo {author} {\bibfnamefont {J.}~\bibnamefont {Herzog-Arbeitman}}, \bibinfo {author} {\bibfnamefont {A.}~\bibnamefont {Chew}}, \bibinfo {author} {\bibfnamefont {K.-E.}\ \bibnamefont {Huhtinen}}, \bibinfo {author} {\bibfnamefont {P.}~\bibnamefont {Törmä}},\ and\ \bibinfo {author} {\bibfnamefont {B.~A.}\ \bibnamefont {Bernevig}},\ }\href {https://doi.org/10.48550/arXiv.2209.00007} {\bibinfo {title} {Many-body superconductivity in topological flat bands}} (\bibinfo {year} {2022}),\ \Eprint {https://arxiv.org/abs/2209.00007} {arXiv:2209.00007 [cond-mat]} \BibitemShut {NoStop}%
\bibitem [{\citenamefont {Liang}\ \emph {et~al.}(2017)\citenamefont {Liang}, \citenamefont {Vanhala}, \citenamefont {Peotta}, \citenamefont {Siro}, \citenamefont {Harju},\ and\ \citenamefont {T{\"o}rm{\"a}}}]{tormaLiangBandGeometryBerry2017}%
  \BibitemOpen
  \bibfield  {author} {\bibinfo {author} {\bibfnamefont {L.}~\bibnamefont {Liang}}, \bibinfo {author} {\bibfnamefont {T.~I.}\ \bibnamefont {Vanhala}}, \bibinfo {author} {\bibfnamefont {S.}~\bibnamefont {Peotta}}, \bibinfo {author} {\bibfnamefont {T.}~\bibnamefont {Siro}}, \bibinfo {author} {\bibfnamefont {A.}~\bibnamefont {Harju}},\ and\ \bibinfo {author} {\bibfnamefont {P.}~\bibnamefont {T{\"o}rm{\"a}}},\ }\bibfield  {title} {\bibinfo {title} {Band geometry, {{Berry}} curvature, and superfluid weight},\ }\href {https://doi.org/10.1103/PhysRevB.95.024515} {\bibfield  {journal} {\bibinfo  {journal} {Physical Review B}\ }\textbf {\bibinfo {volume} {95}},\ \bibinfo {pages} {024515} (\bibinfo {year} {2017})}\BibitemShut {NoStop}%
\bibitem [{\citenamefont {Anderson}(1959)}]{andersonAndersonTheoryDirtySuperconductors1959}%
  \BibitemOpen
  \bibfield  {author} {\bibinfo {author} {\bibfnamefont {P.~W.}\ \bibnamefont {Anderson}},\ }\bibfield  {title} {\bibinfo {title} {Theory of dirty superconductors},\ }\href {https://doi.org/10.1016/0022-3697(59)90036-8} {\bibfield  {journal} {\bibinfo  {journal} {Journal of Physics and Chemistry of Solids}\ }\textbf {\bibinfo {volume} {11}},\ \bibinfo {pages} {26} (\bibinfo {year} {1959})}\BibitemShut {NoStop}%
\bibitem [{\citenamefont {Queiroz}\ \emph {et~al.}(2024)\citenamefont {Queiroz}, \citenamefont {Ilan}, \citenamefont {Song}, \citenamefont {Bernevig},\ and\ \citenamefont {Stern}}]{sternQueirozRingStatesTopological2024}%
  \BibitemOpen
  \bibfield  {author} {\bibinfo {author} {\bibfnamefont {R.}~\bibnamefont {Queiroz}}, \bibinfo {author} {\bibfnamefont {R.}~\bibnamefont {Ilan}}, \bibinfo {author} {\bibfnamefont {Z.}~\bibnamefont {Song}}, \bibinfo {author} {\bibfnamefont {B.~A.}\ \bibnamefont {Bernevig}},\ and\ \bibinfo {author} {\bibfnamefont {A.}~\bibnamefont {Stern}},\ }\href {https://doi.org/10.48550/arXiv.2406.03529} {\bibinfo {title} {Ring states in topological materials}} (\bibinfo {year} {2024}),\ \Eprint {https://arxiv.org/abs/2406.03529} {arXiv:2406.03529 [cond-mat]} \BibitemShut {NoStop}%
\bibitem [{\citenamefont {Julku}\ \emph {et~al.}(2016)\citenamefont {Julku}, \citenamefont {Peotta}, \citenamefont {Vanhala}, \citenamefont {Kim},\ and\ \citenamefont {T{\"o}rm{\"a}}}]{tormaJulkuGeometricOriginSuperfluidity2016}%
  \BibitemOpen
  \bibfield  {author} {\bibinfo {author} {\bibfnamefont {A.}~\bibnamefont {Julku}}, \bibinfo {author} {\bibfnamefont {S.}~\bibnamefont {Peotta}}, \bibinfo {author} {\bibfnamefont {T.~I.}\ \bibnamefont {Vanhala}}, \bibinfo {author} {\bibfnamefont {D.-H.}\ \bibnamefont {Kim}},\ and\ \bibinfo {author} {\bibfnamefont {P.}~\bibnamefont {T{\"o}rm{\"a}}},\ }\bibfield  {title} {\bibinfo {title} {Geometric {{Origin}} of {{Superfluidity}} in the {{Lieb-Lattice Flat Band}}},\ }\href {https://doi.org/10.1103/PhysRevLett.117.045303} {\bibfield  {journal} {\bibinfo  {journal} {Physical Review Letters}\ }\textbf {\bibinfo {volume} {117}},\ \bibinfo {pages} {045303} (\bibinfo {year} {2016})}\BibitemShut {NoStop}%
\bibitem [{Note1()}]{Note1}%
  \BibitemOpen
  \bibinfo {note} {Our assumption is distinct from so-called uniform pairing condition~\cite {huberTovmasyanEffectiveTheoryEmergent2016} which assumes uniform pairing on those orbitals where it is non-zero, and is valid in the Lieb lattice with uniform hoppings.}\BibitemShut {Stop}%
\bibitem [{\citenamefont {Simon}\ and\ \citenamefont {Rudner}(2020)}]{rudnerSimonContrastingLatticeGeometry2020}%
  \BibitemOpen
  \bibfield  {author} {\bibinfo {author} {\bibfnamefont {S.~H.}\ \bibnamefont {Simon}}\ and\ \bibinfo {author} {\bibfnamefont {M.~S.}\ \bibnamefont {Rudner}},\ }\bibfield  {title} {\bibinfo {title} {Contrasting lattice geometry dependent versus independent quantities: {{Ramifications}} for {{Berry}} curvature, energy gaps, and dynamics},\ }\href {https://doi.org/10.1103/PhysRevB.102.165148} {\bibfield  {journal} {\bibinfo  {journal} {Physical Review B}\ }\textbf {\bibinfo {volume} {102}},\ \bibinfo {pages} {165148} (\bibinfo {year} {2020})}\BibitemShut {NoStop}%
\bibitem [{\citenamefont {Peotta}(2022)}]{peottaPeottaSuperconductivityGeneralizedRandom2022}%
  \BibitemOpen
  \bibfield  {author} {\bibinfo {author} {\bibfnamefont {S.}~\bibnamefont {Peotta}},\ }\bibfield  {title} {\bibinfo {title} {Superconductivity, generalized random phase approximation and linear scaling methods},\ }\href {https://doi.org/10.1088/1367-2630/ac9d5c} {\bibfield  {journal} {\bibinfo  {journal} {New Journal of Physics}\ }\textbf {\bibinfo {volume} {24}},\ \bibinfo {pages} {113019} (\bibinfo {year} {2022})}\BibitemShut {NoStop}%
\bibitem [{\citenamefont {Tam}\ and\ \citenamefont {Peotta}(2024)}]{peottaTamGeometryindependentSuperfluidWeight2024}%
  \BibitemOpen
  \bibfield  {author} {\bibinfo {author} {\bibfnamefont {M.}~\bibnamefont {Tam}}\ and\ \bibinfo {author} {\bibfnamefont {S.}~\bibnamefont {Peotta}},\ }\bibfield  {title} {\bibinfo {title} {Geometry-independent superfluid weight in multiorbital lattices from the generalized random phase approximation},\ }\href {https://doi.org/10.1103/PhysRevResearch.6.013256} {\bibfield  {journal} {\bibinfo  {journal} {Physical Review Research}\ }\textbf {\bibinfo {volume} {6}},\ \bibinfo {pages} {013256} (\bibinfo {year} {2024})}\BibitemShut {NoStop}%
\bibitem [{\citenamefont {Moradian}\ \emph {et~al.}(2000)\citenamefont {Moradian}, \citenamefont {Annett}, \citenamefont {Gy{\"o}rffy},\ and\ \citenamefont {Litak}}]{litakMoradianSuperconductingAlloysWeak2000}%
  \BibitemOpen
  \bibfield  {author} {\bibinfo {author} {\bibfnamefont {R.}~\bibnamefont {Moradian}}, \bibinfo {author} {\bibfnamefont {J.~F.}\ \bibnamefont {Annett}}, \bibinfo {author} {\bibfnamefont {B.~L.}\ \bibnamefont {Gy{\"o}rffy}},\ and\ \bibinfo {author} {\bibfnamefont {G.}~\bibnamefont {Litak}},\ }\bibfield  {title} {\bibinfo {title} {Superconducting alloys with weak and strong scattering: {{Anderson}}'s theorem and a superconductor-insulator transition},\ }\href {https://doi.org/10.1103/PhysRevB.63.024501} {\bibfield  {journal} {\bibinfo  {journal} {Physical Review B}\ }\textbf {\bibinfo {volume} {63}},\ \bibinfo {pages} {024501} (\bibinfo {year} {2000})}\BibitemShut {NoStop}%
\bibitem [{\citenamefont {Murthy}\ and\ \citenamefont {Shankar}(2003)}]{shankarMurthyHamiltonianTheoriesFractional2003}%
  \BibitemOpen
  \bibfield  {author} {\bibinfo {author} {\bibfnamefont {G.}~\bibnamefont {Murthy}}\ and\ \bibinfo {author} {\bibfnamefont {R.}~\bibnamefont {Shankar}},\ }\bibfield  {title} {\bibinfo {title} {Hamiltonian theories of the fractional quantum hall effect},\ }\href {https://doi.org/10.1103/RevModPhys.75.1101} {\bibfield  {journal} {\bibinfo  {journal} {Reviews of Modern Physics}\ }\textbf {\bibinfo {volume} {75}},\ \bibinfo {pages} {1101} (\bibinfo {year} {2003})}\BibitemShut {NoStop}%
\bibitem [{\citenamefont {Claassen}\ \emph {et~al.}(2015)\citenamefont {Claassen}, \citenamefont {Lee}, \citenamefont {Thomale}, \citenamefont {Qi},\ and\ \citenamefont {Devereaux}}]{devereauxClaassenPositionMomentumDualityFractional2015}%
  \BibitemOpen
  \bibfield  {author} {\bibinfo {author} {\bibfnamefont {M.}~\bibnamefont {Claassen}}, \bibinfo {author} {\bibfnamefont {C.~H.}\ \bibnamefont {Lee}}, \bibinfo {author} {\bibfnamefont {R.}~\bibnamefont {Thomale}}, \bibinfo {author} {\bibfnamefont {X.-L.}\ \bibnamefont {Qi}},\ and\ \bibinfo {author} {\bibfnamefont {T.~P.}\ \bibnamefont {Devereaux}},\ }\bibfield  {title} {\bibinfo {title} {Position-momentum duality and fractional quantum hall effect in chern insulators},\ }\href {https://doi.org/10.1103/PhysRevLett.114.236802} {\bibfield  {journal} {\bibinfo  {journal} {Physical Review Letters}\ }\textbf {\bibinfo {volume} {114}},\ \bibinfo {pages} {236802} (\bibinfo {year} {2015})}\BibitemShut {NoStop}%
\bibitem [{\citenamefont {Li}\ \emph {et~al.}(2024)\citenamefont {Li}, \citenamefont {Dong}, \citenamefont {Ledwith},\ and\ \citenamefont {Khalaf}}]{khalafLiConstraintsRealSpace2024}%
  \BibitemOpen
  \bibfield  {author} {\bibinfo {author} {\bibfnamefont {Q.}~\bibnamefont {Li}}, \bibinfo {author} {\bibfnamefont {J.}~\bibnamefont {Dong}}, \bibinfo {author} {\bibfnamefont {P.~J.}\ \bibnamefont {Ledwith}},\ and\ \bibinfo {author} {\bibfnamefont {E.}~\bibnamefont {Khalaf}},\ }\href@noop {} {\bibinfo {title} {Constraints on real space representations of chern bands}} (\bibinfo {year} {2024}),\ \Eprint {https://arxiv.org/abs/2407.02561} {arXiv:2407.02561 [cond-mat]} \BibitemShut {NoStop}%
\bibitem [{\citenamefont {Roy}(2014)}]{royRoyBandGeometryFractional2014}%
  \BibitemOpen
  \bibfield  {author} {\bibinfo {author} {\bibfnamefont {R.}~\bibnamefont {Roy}},\ }\bibfield  {title} {\bibinfo {title} {Band geometry of fractional topological insulators},\ }\href {https://doi.org/10.1103/PhysRevB.90.165139} {\bibfield  {journal} {\bibinfo  {journal} {Physical Review B}\ }\textbf {\bibinfo {volume} {90}},\ \bibinfo {pages} {165139} (\bibinfo {year} {2014})}\BibitemShut {NoStop}%
\bibitem [{\citenamefont {Wang}\ \emph {et~al.}(2021)\citenamefont {Wang}, \citenamefont {Cano}, \citenamefont {Millis}, \citenamefont {Liu},\ and\ \citenamefont {Yang}}]{yangWangExactLandauLevel2021}%
  \BibitemOpen
  \bibfield  {author} {\bibinfo {author} {\bibfnamefont {J.}~\bibnamefont {Wang}}, \bibinfo {author} {\bibfnamefont {J.}~\bibnamefont {Cano}}, \bibinfo {author} {\bibfnamefont {A.~J.}\ \bibnamefont {Millis}}, \bibinfo {author} {\bibfnamefont {Z.}~\bibnamefont {Liu}},\ and\ \bibinfo {author} {\bibfnamefont {B.}~\bibnamefont {Yang}},\ }\bibfield  {title} {\bibinfo {title} {Exact landau level description of geometry and interaction in a flatband},\ }\href {https://doi.org/10.1103/PhysRevLett.127.246403} {\bibfield  {journal} {\bibinfo  {journal} {Physical Review Letters}\ }\textbf {\bibinfo {volume} {127}},\ \bibinfo {pages} {246403} (\bibinfo {year} {2021})},\ \Eprint {https://arxiv.org/abs/2105.07491} {arXiv:2105.07491 [cond-mat]} \BibitemShut {NoStop}%
\bibitem [{\citenamefont {Arora}\ \emph {et~al.}(2025)\citenamefont {Arora}, \citenamefont {Curtis},\ and\ \citenamefont {Narang}}]{narangAroraQuantumGeometryInduced2025}%
  \BibitemOpen
  \bibfield  {author} {\bibinfo {author} {\bibfnamefont {A.}~\bibnamefont {Arora}}, \bibinfo {author} {\bibfnamefont {J.~B.}\ \bibnamefont {Curtis}},\ and\ \bibinfo {author} {\bibfnamefont {P.}~\bibnamefont {Narang}},\ }\bibfield  {title} {\bibinfo {title} {Quantum geometry induced microwave enhancement of superconducting order in flat bands},\ }\href {https://doi.org/10.1038/s42005-025-02244-5} {\bibfield  {journal} {\bibinfo  {journal} {Communications Physics}\ }\textbf {\bibinfo {volume} {8}},\ \bibinfo {pages} {327} (\bibinfo {year} {2025})}\BibitemShut {NoStop}%
\bibitem [{\citenamefont {Marzari}\ \emph {et~al.}(2012)\citenamefont {Marzari}, \citenamefont {Mostofi}, \citenamefont {Yates}, \citenamefont {Souza},\ and\ \citenamefont {Vanderbilt}}]{vanderbiltMarzariMaximallyLocalizedWannier2012}%
  \BibitemOpen
  \bibfield  {author} {\bibinfo {author} {\bibfnamefont {N.}~\bibnamefont {Marzari}}, \bibinfo {author} {\bibfnamefont {A.~A.}\ \bibnamefont {Mostofi}}, \bibinfo {author} {\bibfnamefont {J.~R.}\ \bibnamefont {Yates}}, \bibinfo {author} {\bibfnamefont {I.}~\bibnamefont {Souza}},\ and\ \bibinfo {author} {\bibfnamefont {D.}~\bibnamefont {Vanderbilt}},\ }\bibfield  {title} {\bibinfo {title} {Maximally localized {{Wannier}} functions: {{Theory}} and applications},\ }\href {https://doi.org/10.1103/RevModPhys.84.1419} {\bibfield  {journal} {\bibinfo  {journal} {Reviews of Modern Physics}\ }\textbf {\bibinfo {volume} {84}},\ \bibinfo {pages} {1419} (\bibinfo {year} {2012})}\BibitemShut {NoStop}%
\bibitem [{\citenamefont {Canc{\`e}s}\ and\ \citenamefont {Le~Bris}(2000)}]{lebrisCancesCanWeOutperform2000}%
  \BibitemOpen
  \bibfield  {author} {\bibinfo {author} {\bibfnamefont {E.}~\bibnamefont {Canc{\`e}s}}\ and\ \bibinfo {author} {\bibfnamefont {C.}~\bibnamefont {Le~Bris}},\ }\bibfield  {title} {\bibinfo {title} {Can we outperform the {{DIIS}} approach for electronic structure calculations?},\ }\href {https://doi.org/10.1002/1097-461X(2000)79:2<82::AID-QUA3>3.0.CO;2-I} {\bibfield  {journal} {\bibinfo  {journal} {International Journal of Quantum Chemistry}\ }\textbf {\bibinfo {volume} {79}},\ \bibinfo {pages} {82} (\bibinfo {year} {2000})}\BibitemShut {NoStop}%
\end{thebibliography}%
